\documentclass[english,11pt,a4paper,preprintnumbers,aps,prd,nofootinbib,showpacs,superscriptaddress, notitlepage]{revtex4-2} 

\usepackage{color}
\usepackage{float}
\usepackage{graphicx}
\usepackage{multirow}
\usepackage{amsmath}
\usepackage{amssymb}
\usepackage[colorlinks=true,citecolor=darkred,urlcolor=darkred, pdfborder={0 0 0}]{hyperref}
\usepackage[normalem]{ulem}
\definecolor{darkred}{rgb}{0.6,0,0}
\definecolor{drkgrn}{RGB}{0, 51, 0}
\usepackage{soul}

\definecolor{linkcolor}{rgb}{0,0,0.5}

\def\gsim{\raise0.3ex\hbox{$\;>$\kern-0.75em\raise-1.1ex\hbox{$\sim\;$}}}
\def\lsim{\raise0.3ex\hbox{$\;<$\kern-0.75em\raise-1.1ex\hbox{$\sim\;$}}}

\def\beqn#1{\begin{equation}\label{#1}}
\def\eeqn{\end{equation}}

\def\beqa#1{\begin{eqnarray}\label{#1}}
\def\eeqa{\end{eqnarray}}

\def\Z2{$\mathcal{Z_2}$}

\newcommand {\ignore}[1]{}

\usepackage[T1]{fontenc} % if needed
\usepackage{graphicx}
\usepackage{mathtools}
\usepackage{tikz-feynman}

\mathchardef\mhyphen="2D % Define a "math hyphen"

\newcommand{\h}{\langle H \rangle}
\newcommand{\hu}{\langle \widetilde{H} \rangle}
\newcommand{\s}{\langle S \rangle}
\newcommand{\fn}{\rm{FN}}

\newcommand{\cw}{\rm{CW}}

\begin{document}

\title{\boldmath Probing intermediate scale Froggatt-Nielsen models at future gravitational wave observatories}
%\title{\boldmath Peccei-Quinn as a Froggatt Nielsen symmetry for the quark and lepton masses.}
%Flaxion: Unifying flavor symmetry with Peccei-Quinn
\author{Dhruv Ringe}\email{phd1901151004@iiti.ac.in}\affiliation{Discipline of Physics, IIT Indore,\\Khandwa Road, Simrol, Indore, Madhya Pradesh, India -453552}

\vspace{2.0cm}
%\center{\bf Abstract}
\begin{abstract}
{\noindent The flavor symmetry-breaking scale in the Froggatt-Nielsen (FN) mechanism is very weakly constrained by present experiments, and can lie anywhere from a few TeV to the Planck scale. We construct two minimal, non-supersymmetric, ultraviolet (UV) complete models that generate the FN mechanism, with a global $U(1)_{\fn}$ flavor symmetry and a single flavon field. Using the one-loop finite temperature effective potential, we explore the possibility of a strong first order phase transition (SFOPT) induced by the flavon. We show that if the flavor symmetry-breaking occurs at intermediate scales  $\sim 10^4-10^7$ GeV, then in certain regions of the parameter space, the associated stochastic gravitational wave (GW) background is strong enough to be detected by second generation GW observatories such as the Big Bang Observer (BBO), the Deci-hertz interferometer Gravitational Observatory (DECIGO), the Cosmic Explorer (CE) and the Einstein Telescope (ET). We identify viable regions of the parameter space for the best detection prospects. While both models of flavor can produce a detectable GW background, the GW signature cannot be used to discriminate between them.}
\end{abstract}

\maketitle
\newpage
\section{Introduction}
\label{sec:intro}

The observation of the first GW signal from a binary neutron star merger GW2015 \cite{Abbott_2016} by the aLIGO \cite{Harry:2010zz} collaboration has opened up a new window to explore the early history of the universe. Our capability of observing feeble GW signals is expected to be significantly enhanced by several planned GW observatories such as LISA \cite{LISA:2017pwj}, BBO \cite{Corbin_2006}, DECIGO \cite{Musha:2017usi}, CE \cite{Abbott_2017}, and ET \cite{Punturo:2010zz}. Along with pulsar timing arrays (PTA's) \citep{NANOGrav:2020bcs,Manchester:2013ndt,Burke-Spolaor:2018bvk,Hobbs:2009yy,Lentati:2015qwp,Manchester:2012za}, these observatories will detect GW signals with frequencies ranging from $\sim 10^{-9}$ Hz to $\sim 10^4$ Hz. With binary neurton star and binary black hole merger events routinely being detected by LIGO and VIRGO collaborations \citep{Acernese_2014,Abbott_2019,Abbott_2020}, the prospect of detecting a stochastic cosmological background of GW's by these new obervatories is indeed promising. Several possible sources of a stochastic GW background have been proposed, including the density fluctuations from inflation \citep{Matarrese:1997ay,Mollerach:2003nq,Baumann:2007zm,Kohri:2018awv,DEramo:2019tit}, topological defects in the vacuum \citep{Vilenkin:1984ib,Caldwell:1991jj,Hindmarsh:1994re,Preskill:1991kd,Gleiser:1998na,Hiramatsu:2013qaa}, and cosmological phase transitions \citep{Caprini:2015zlo,Caprini:2018mtu}. 

The final piece of the minimal Standard Model (SM) of particle physics was found with the discovery of the Higgs boson  at the Large Hadron Collider (LHC) in 2012 \citep{ATLAS:2012yve,CMS:2012qbp}. Despite its tremendous success, the SM has limitations which hint at possible new physics. The SM does not explain the hierarchy of fermion masses, or equivalently the hierarchy of yukawa couplings, which range from $Y\sim 1$ for the top-quark, to $Y\sim 10^{-6}$ for the electron. Accounting for neutrino masses further enhances this hierarchy, making the problem much worse. Equally puzzling is the hierarchy of quark mixing angles, and the anarchy of leptonic mixing angles. This problem of yukawa hierarchies has come to be known as the SM flavor puzzle \citep{Babu:2009fd,Grossman:2017thq,Nir:2020jtr}.

In this paper, we revisit the elegant solution to the SM flavor puzzle, originially proposed by Froggatt and Nielsen \cite{Froggatt:1978nt}, focussing on the quark sector. The observed hiearchy of quark masses and mixings  is generated by non-renormalizable operators involving  one or more complex scalar fields (flavons). The flavons spontaneously break the flavor symmetry imposed on the SM fields by acquiring a non-zero vacuum expectation value (VEV). A crucial parameter in FN models is the ratio $\epsilon = \frac{\langle S \rangle}{\Lambda}$, where $\langle S \rangle$ is the VEV of the flavon, and $\Lambda$ is the cutoff scale of the effective theory, associated with the mass of the messenger vector-like fermions (VLF's) in the UV completion. 

The UV completion of FN is a tedious task due to the involvement of several VLF's, and is further complicated for realizations of FN with large $U(1)_{\fn}$ charge assignments. Some explicit UV constructions are found for example in \citep{Nir-Seiberg1,Nir-Seiberg2,Calibbi2012yj,delaVega:2021ugs}. While the method to determine the minimum number vectorlike fermions was first given in \cite{Nir-Seiberg1}, a simple procedure was given in \cite{Calibbi2012yj} to find the complete set of charge assignments. Using the procedure of \cite{Calibbi2012yj}, we construct minimal, non-supersymmetric, UV complete models for two simple realizations of the FN mechanism with one flavon. It should be noted however that adding several heavy fermions threatens the vacuum stability of the flavon potential, which demands the inclusion of additional bosons to restore the vacuum stability. 

Given an FN model, the ratio $\epsilon$ is fixed to some extent by the observed pattern of quark masses and mixing angles, whereas the scale $\Lambda$ itself is undetermined. Phenomenology of neutral meson mixing puts a lower bound of a few TeV on $\Lambda$ \cite{huntingflavon}. On the other hand, searches for vector-like fermions based on collider results \citep{Ellis:2014dza,Buchkremer:2012dn} and electroweak vacuum stability \citep{Blum:2015rpa,Gopalakrishna:2018uxn,Borah:2020nsz} put a bound of $\gtrsim1$ TeV. Planned future colliders such as the International Linear Collider (ILC) \citep{Behnke:2013xla,Baer:2013cma} and the Future Circular Collider (FCC) \citep{FCC:2018byv,FCC:2018evy,FCC:2018vvp} may push the lower bound on $\Lambda$ to $\sim 50$ TeV. Beyond that, the cutoff is practically unconstrained and could lie anywhere between $\sim 10$ TeV and the Planck scale $(\sim 10^{19}$ GeV). However, if the flavon undergoes a SFOPT, then the corresponding stochastic background may be detected by upcoming GW detectors, for symmetry-breaking scales as high as $10^4-10^7$ GeV. The corresponding GW signature has a peak frequency of $10^{-1}-10^2$ Hz, making it an ideal energy scale to be probed at observatiories like BBO, DECIGO, ET and CE. While several authors have discussed the GW background arising at such intermediate scales in different contexts \citep{Dev:2016feu,Addazi:2018nzm,Croon:2018kqn,Baldes:2018nel,Dev:2019njv,Greljo:2019xan,VonHarling:2019rgb,Huang:2020bbe,Chu:2022mjw}, as far as we know, this is the first attempt to probe FN models incorporating UV completions, with GW's.

This paper is organized as follows: in section \ref{sec:FN}, we review the FN mechanism and discuss two realizations involving a single flavon. In section \ref{sec: UV}, we construct the UV theory for the two FN models by giving the particle content involving heavy vector-like quarks. In section \ref{sec: potential}, we construct the one-loop finite temperature effective  potential needed to analyse the flavon phase transition (PT). The issue of vacuum stability of FN UV completions is discussed in section \ref{sec: stability}. In section \ref{sec:FOPT} we define the parameters characterizing a first order phase transition (FOPT), and give the parameter scans showing regions of SFOPT in the two models considered. In section \ref{sec:GW}, we discuss the prospects of dectecting the GW signature at the future GW observatories. Finally, we give our concluding remarks in section \ref{sec: conclusion}.

\section{Froggatt-Nielsen mechanism}
\label{sec:FN}
In its simplest form, the FN mechanism is achieved by imposing a global horizontal $U(1)_{\fn}$ symmetry on the SM fields. This symmetry prohibits the inclusion of the usual yukawa terms in the SM Lagrangian. Instead, we can write $U(1)_{\fn}$ preserving effective operators of the form:
\begin{equation}
  - \mathcal{L}_{\fn} = y_{ij}\left(\frac{S}{\Lambda}\right)^{n_{ij}} \overline{f}_{Li} H f_{Rj} + \rm{h.c.} 
\end{equation}
where $y_{ij}$ are $\mathcal{O}(1)$ couplings, $f_L$, $f_R$ respectively represent left handed fermion doublets and right handed fermion singlets of the SM, and $i,j$ denote the family index. $\Lambda$ is the cutoff scale, related to the mass scale of particles of the UV theory. $S$ is a complex scalar field called the flavon, and $H$ is the SM Higgs doublet. The FN mechanism works both in the lepton sector as well as the quark sector. Here we focus exlusively on the quark sector, where,
\begin{equation}\label{eq:FN}
-\mathcal{L}_{\fn} = y^u_{ij}\left(\frac{S}{\Lambda}\right)^{n^u_{ij}} \overline{Q}^i_{L} \widetilde{H}~u^j_{R} + y^d_{ij}\left(\frac{S}{\Lambda}\right)^{n^d_{ij}} \overline{Q}^i_{L} H~ d^{~j}_{R} + \rm{h.c.}
\end{equation} 
where $\widetilde{H} = i \sigma^2 H$. Without loss of generality, we can choose the FN charges of the scalars as: $Q_{\fn}(S) = -1$ and $Q_{\fn}(H) = 0$. The power of the $\frac{S}{\Lambda}$ term is: $$n^{u/d}_{ij} = Q_{\fn}(\overline{Q}_L^i)+Q_{\fn}(u_R^{~j}/d_R^{~j}).$$

When $U(1)_{\fn}$ is spontaneously broken by the non-zero VEV of the flavon, the FN operators give rise to the familiar SM yukawa terms:
\begin{equation} \label{eq: yukSM}
-\mathcal{L}_{\rm{Y}} = Y^u_{ij}\overline{Q}^i_{L} \widetilde{H}~u^j_{R} + Y^d_{ij} \overline{Q}^i_{L} H~ d^{~j}_{R} + \rm{h.c.},
\end{equation}
where ,
\begin{equation}\label{eq: yij}
Y^{u/d}_{ij} = y^{u/d}_{ij} \epsilon^{n^{u/d}_{ij}},
\end{equation}
and $\epsilon$ is defined as the ratio,
\begin{equation}
\epsilon \equiv \frac{\langle S \rangle}{\Lambda} = \frac{v_s}{\sqrt{2}\Lambda}
\end{equation}
For $\epsilon<1$, the desired hierarchy of quark masses is generated with appropriate charge assigments. Corresponding quark mass matrices are: $${\bf{\mathcal{M}^{u/d}}} = \frac{v}{\sqrt{2}}{\bf{Y^{u/d}}},$$ where $\frac{v}{\sqrt{2}}$  is the electroweak VEV with $v = 246.2$ GeV.

It is useful to parameterize the hierarchy in terms of the Cabbibo angle: $\lambda=\sin\theta_{cb}\sim 0.23$. The effective couplings $y^{u/d}_{ij}$ are all required to be $\mathcal{O}(1)$, but their exact value is unimportant, allowing some freedom in assigning charges to the quarks.

Below we consider two simple examples which implement the FN mechanism with a single flavon. 

\begin{itemize}
\item \textbf{Model 1}: $\epsilon = \lambda = 0.23$, \cite{Binetruy:1996xk}
\begin{equation}\label{eq: model1}
Q_{\fn}(\overline{Q}_L)=(3,2,0),~Q_{\fn}(u_R)=(5,2,0),~Q_{\fn}(d_R)=(4,3,3).
\end{equation}
The yukawa matrices are:
\begin{gather}\label{eq: yuk_model1}
{\bf{Y^u}} \sim \begin{pmatrix}
\epsilon^8 & \epsilon^5 & \epsilon^3\\
\epsilon^7 & \epsilon^4 & \epsilon^2\\
\epsilon^5 & \epsilon^2 & 1
\end{pmatrix},~~~
{\bf{Y^d}} \sim \begin{pmatrix}
\epsilon^7 & \epsilon^6 & \epsilon^6\\
\epsilon^6 & \epsilon^5 & \epsilon^5\\
\epsilon^4 & \epsilon^3 & \epsilon^3
\end{pmatrix}, 
\end{gather}

where each entry of the matrices is multiplied by an $\mathcal{O}(1)$ factor $y^{u/d}_{ij}$. It is useful to note the $\epsilon$-dependence of the determinants:

\begin{equation}\label{det1}
\det {\bf{Y^u}} \propto \epsilon^{12},~~~\det {\bf{Y^d}} \propto \epsilon^{15}.
\end{equation}

\item \textbf{Model 2}: This is a non-supersymmetric variation of the charge assignment considered in \cite{Berkooz:2004kx}. $\epsilon = \lambda^2 = 0.05$, 
\begin{equation}\label{eq: model2}
Q_{\fn}(\overline{Q}_L)= Q_{\fn}(u_R)=(2,1,0),~Q_{\fn}(d_R)=(1,1,1).
\end{equation}
The yukawa matrices are \footnote{Note that the matrix $\bf{Y^d}$ differs from the matrix in \cite{Berkooz:2004kx} by a factor of $\epsilon$. This is because of the presence of two Higgs doublets in supersymmetric theories, which can take care of the additional hierarchy.}:
\begin{gather}\label{eq: yuk_model2}
{\bf{Y^u}} \sim \begin{pmatrix}
\epsilon^4 & \epsilon^3 & \epsilon^2\\
\epsilon^3& \epsilon^2 & \epsilon\\
\epsilon^2 & \epsilon & 1
\end{pmatrix},~~~
{\bf{Y^d}} \sim \begin{pmatrix}
\epsilon^3 & \epsilon^3 & \epsilon^3\\
\epsilon^2 & \epsilon^2 & \epsilon^2\\
\epsilon & \epsilon & \epsilon
\end{pmatrix}.
\end{gather}
The determinants are:
\begin{equation}\label{det2}
\det {\bf{Y^u}} \propto \epsilon^6,~~~\det {\bf{Y^d}} \propto \epsilon^6.
\end{equation}
\end{itemize}

The FN mechanism can also be realized with more complicated abelian/non-abelian groups instead of $U(1)_{\fn}$, that may or may not be gauged. As a result, there can be models with more than one flavon field. See Appendix \ref{app: model3} for an example of a two flavon model. As we'll see in the next section, lower powers of $\epsilon$ imply the need for fewer heavy quarks, and hence a simpler UV theory. 

\section{UV completion of FN models}\label{sec: UV}

\begin{figure}[tbp]
\centering % \begin{center}/\end{center} takes some additional vertical space
\includegraphics[width=.95\textwidth]{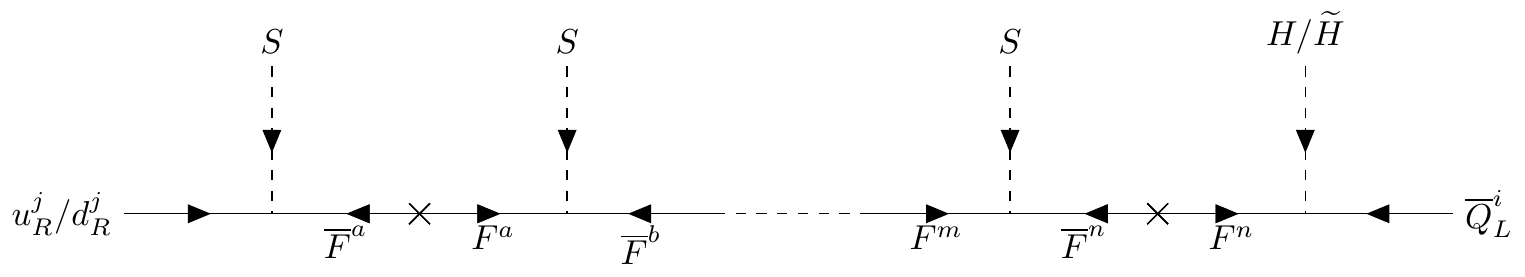}
\hfill
\caption{\label{fig: chain} A chain diagram of the UV theory, contributing to an FN operator corresponding to $Y^{u/d}_{ij}$. The number of insertions of the $S$ field is equal to the power of $\epsilon$ in $Y^{u/d}_{ij}$. Here $F$'s represent the heavy VLQ's sitting at the scale $\Lambda$. Yukawa couplings at the vertices are all $\mathcal{O}(1)$. Arrows indicate the flow of $U(1)_{\fn}$ charge, which is conserved at each vertex.}
\end{figure}

In the high energy theory, the effective operators listed in \eqref{eq:FN}. are generated from chain diagrams of the kind shown in figure \ref{fig: chain}. These diagrams are mediated by a set of heavy vector-like quarks (VLQ's) with mass $M\sim \Lambda$. We assume all VLQ's to have the same bare mass $M$. The heavy VLQ's can either be $SU(2)_L$ singlets, or doublets. For simplicity we take all the VLQ's to be $SU(2)_L$ singlets, as it does not impact our ultimate goal of analysing the flavon phase transition. We represent the chain diagram of figure \ref{fig: chain} using the short hand notation,
\begin{equation*}
u^j/d^j-F^a-F^b-\cdots -F^m-F^n-\overline{Q}^i
\end{equation*}
The scalar field at each vertex is suppressed, and only the SM fermions and the mediating VLQ's are shown. For $SU(2)_L$ singlet VLQ's, it is assumed that the scalar attached to $\overline{Q}$ is $H/\widetilde{H}$, while an $S$ field is attached to each of the other vertices.

The singlet VLQ's come in `up-type' or `down-type' representations of the SM gauge group, $\mathcal{G}_{\rm{SM}}=SU(3)_c\times SU(2)_L\times U_1(Y),$ depending upon the type of light quark they interact with. The notation used here is: 
$$(T_L, T_R)\sim ({\bf{3}, \bf{1}}, 4/3)~~({\rm{up\mhyphen type}}),$$
$$(B_L, B_R)\sim ({\bf{3}, \bf{1}},-2/3)~~({\rm{down\mhyphen type}}).$$ 
The fermionic mass Lagrangian can be written as,
\begin{equation}
-\mathcal{L}_{\rm{UV}} \supset \overline{F^u}_L \mathcal{M}^u F^u_R + \overline{F^d}_L \mathcal{M}^d F^d_R + \rm{h.c.},
\end{equation}
where $F^u_L = \big(Q^1,Q^2,Q^3, T^1,T^2\cdots\big)_L^T$, 
$F^u_R = \big(u,c,t, T^1,T^2\cdots\big)_R^T$, 
$F^d_L = \big(Q^1,Q^2,Q^3, B^1,B^2\cdots\big)_L^T$, $F^d_L = \big(d,s,b, B^1,B^2\cdots\big)_R^T$. 

The mass matrices $\mathcal{M}^{u/d}$ depend on the specific model. Physical fermion masses are given by the singular values of $\mathcal{M}^{u/d}$, i.e. eigenvalues of $\big(\mathcal{M}^{\dagger}\mathcal{M}\big)^{1/2}$. In \citep{Nir-Seiberg1,Nir-Seiberg2}, it was shown that for a yukawa matrix $\textbf{Y}$ of the SM fermions with $\det\textbf{Y}\propto \epsilon^n$, at least $n$ VLQ's are required in the UV theory. We now follow the procedure elucidated in \cite{Calibbi2012yj} to assign charges to the VLQ's, and build the particle content for our UV models.  

\subsection{Model 1}
According to \eqref{det1}, atleast $12$ u-type VLQ's and $15$ d-type VLQ's are needed. For the up sector, first we select three elements from the $\textbf{Y}^u$ matrix in \eqref{eq: yuk_model1}, which contribute to the determinant \footnote{Every term in the determinant comes with the same power of $\epsilon$. This is a general feature of choosing a $U(1)$ symmetry.}. For each element, we form a chain diagram with the number of VLQ's equal to the corresponding power of $\epsilon$. Here we choose, $Y^u_{11}\propto \epsilon^8$, $Y^u_{22}\propto \epsilon^4$, and $Y^u_{33}\propto \epsilon^0$. Next, we use a separate set of VLQ's to construct each chain. The $U(1)_{\fn}$ charge assignment is determined by conserving the charge at at every vertex.
\begin{eqnarray}
&u_R^1-T^8-T^7-T^6-T^5-T^4-T^3-T^2-T^1-\overline{Q}_L^1& \nonumber \\
&u_R^2-T'^4-T'^3-T'^2-T'^1-\overline{Q}_L^2& \nonumber
\end{eqnarray}

\noindent Similarly, for the down sector, we have: $Y^d_{11}\propto \epsilon^7$, $Y^d_{22}\propto \epsilon^5$, and $Y^d_{33}\propto \epsilon^3$. The chains are, 
\begin{eqnarray}
&d_R^1-B^7-B^6-B^5-B^4-B^3-B^2-B^1-\overline{Q}_L^1& \nonumber \\
&d_R^2-B'^5-B'^4-B'^3-B'^2-B'^1-\overline{Q}_L^2& \nonumber \\
&d_R^3-B''^3-B''^2-B''^1-\overline{Q}_L^3& \nonumber
\end{eqnarray}

\begin{table}[tbp]
\small
\begin{center}
\vspace{0.5 cm}
\begin{tabular}{|c|c c c c c c c c|}
\hline
 $Q_{\fn}$ & -3    & -2     & -1      & 0      & 1      & 2      & 3       & 4    \\ 
\hline
up-type    & $T^1$ & $T^2$  & $T^3$   & $T^4$  & $T^5$  & $T^6$  & $T^7$   & $T^8$\\  
           &       & $T'^1$ & $T'^2$  & $T'^3$ & $T'^4$ &        &         &      \\    
\hline
down-type  & $B^1$ & $B^2$  & $B^3$   & $B^4$  & $B^5$  & $B^6$  & $B^7$   &      \\  
           &       & $B'^1$ & $B'^2$  & $B'^3$ & $B'^4$ & $B'^5$ &         &      \\
           &       &        &         & $B''^1$& $B''^2$& $B''^3$&         &      \\
\hline
\end{tabular}
\end{center}
\caption{$U(1)_{\fn}$ charges for VLQ's in model 1. Total $12+15=27$ heavy quarks are needed in this model. \label{table:table1}
}
\end{table}

\noindent Chains for the rest of the terms can now be written using the VLQ's found above. The $U(1)_{\fn}$ charge assignments for the VLQ's are displayed in table \ref{table:table1}. The resulting mass matrices are,

\begin{subequations}\label{subeq:model1}
\begin{equation}
\mathcal{M}^u = \begin{psmallmatrix}
0& 0& 0& \hu& 0& 0& 0& 0& 0 & 0 & 0 & 0 & 0 & 0 & 0\\
0& 0& 0& 0& \hu& \hu& 0& 0& 0 & 0 & 0 & 0 & 0 & 0 & 0\\
0& 0& \hu& 0& 0& 0& 0& 0& \hu & \hu & 0 & 0 & 0 & 0 & 0\\
0& 0& 0& M& \s& \s& 0& 0& 0 & 0 & 0 & 0 & 0 & 0 & 0\\
0& 0& 0& 0& M& 0& \s& \s& 0& 0 & 0 & 0 & 0 & 0 & 0\\
0& 0& 0& 0& 0& M& \s& \s& 0& 0 & 0 & 0 & 0 & 0 & 0\\
0& 0& \s& 0& 0& 0& M& 0& \s& \s & 0 & 0 & 0 & 0 & 0\\
0& 0& \s& 0& 0& 0& 0& M& \s& \s & 0 & 0 & 0 & 0 & 0\\
0& 0& 0& 0& 0& 0& 0& 0& M& 0 & \s & \s & 0 & 0 & 0\\
0& 0& 0& 0& 0& 0& 0& 0& 0& M & \s & \s & 0 & 0 & 0\\
0& \s& 0& 0& 0& 0& 0& 0& 0& 0 & M & 0 & \s & 0 & 0\\
0& \s& 0& 0& 0& 0& 0& 0& 0& 0 & 0 & M & \s & 0 & 0\\
0& 0& 0& 0& 0& 0& 0& 0& 0& 0 & 0 & 0 & M & \s & 0\\
0& 0& 0& 0& 0& 0& 0& 0& 0& 0 & 0 & 0 & 0 & M & \s\\
\s& 0& 0& 0& 0& 0& 0& 0& 0& 0 & 0 & 0 & 0 & 0 & M\\
\end{psmallmatrix},
\end{equation}
\begin{equation}
\mathcal{M}^d = \begin{psmallmatrix}
0& 0& 0& \h& 0& 0& 0& 0& 0& 0& 0& 0& 0& 0& 0& 0& 0& 0\\
0& 0& 0& 0& \h& \h& 0& 0& 0& 0& 0& 0& 0& 0& 0& 0& 0& 0\\
0& 0& 0& 0& 0& 0& 0& 0& \h& \h& \h& 0& 0& 0& 0& 0& 0& 0\\
0& 0& 0& M& \s& \s& 0& 0& 0& 0& 0& 0& 0& 0& 0& 0& 0& 0\\
0& 0& 0& 0& M& 0& \s& \s& 0& 0& 0& 0& 0& 0& 0& 0& 0& 0\\
0& 0& 0& 0& 0& M& \s& \s& 0& 0& 0& 0& 0& 0& 0& 0& 0& 0\\
0& 0& 0& 0& 0& 0& M& 0& \s& \s& \s& 0& 0& 0& 0& 0& 0& 0\\
0& 0& 0& 0& 0& 0& 0& M& \s& \s& \s& 0& 0& 0& 0& 0& 0& 0\\
0& 0& 0& 0& 0& 0& 0& 0& M& 0& 0& \s& \s& \s& 0& 0& 0& 0\\
0& 0& 0& 0& 0& 0& 0& 0& 0& M& 0& \s& \s& \s& 0& 0& 0& 0\\
0& 0& 0& 0& 0& 0& 0& 0& 0& 0& M& \s& \s& \s& 0& 0& 0& 0\\
0& 0& 0& 0& 0& 0& 0& 0& 0& 0& 0& M& 0& 0& \s& \s& \s& 0\\
0& 0& 0& 0& 0& 0& 0& 0& 0& 0& 0& 0& M& 0& \s& \s& \s& 0\\
0& 0& 0& 0& 0& 0& 0& 0& 0& 0& 0& 0& 0& M& \s& \s& \s& 0\\
0& \s& \s& 0& 0& 0& 0& 0& 0& 0& 0& 0& 0& 0& M& 0& 0& \s\\
0& \s& \s& 0& 0& 0& 0& 0& 0& 0& 0& 0& 0& 0& 0& M& 0& \s\\
0& \s& \s& 0& 0& 0& 0& 0& 0& 0& 0& 0& 0& 0& 0& 0& M& \s\\
\s& 0& 0& 0& 0& 0& 0& 0& 0& 0& 0& 0& 0& 0& 0& 0& 0& M
\end{psmallmatrix}.
\end{equation}
\end{subequations}
The first three rows and columns of each matrix correspond to the respective light quarks in UV theory. The VLQ's are arranged in order of increasing $Q_{\fn}$. 

\subsection{Model 2}
According to \eqref{det2}, atleast $6$ u-type VLQ's and $6$ d-type VLQ's are needed. Repeating the procedure described above, for the up sector we have: $Y^u_{11}\propto \epsilon^4$, $Y^u_{22}\propto \epsilon^2$, and $Y^u_{33}\propto\epsilon^0$. Therefore we have the chains,
\begin{eqnarray}
&u_R^1-T^4-T^3-T^2-T^1-\overline{Q}_L^1& \nonumber \\
&u_R^2-T'^2-T'^1-\overline{Q}_L^2& \nonumber
\end{eqnarray}

\noindent For the down sector, we have: $Y^d_{11}\propto \epsilon^3$, $Y^d_{22}\propto\epsilon^2$, and $Y^d_{33}\propto\epsilon^1$. This gives,
\begin{eqnarray}
&d_R^1-B^3-B^2-B^1-\overline{Q}_L^1& \nonumber \\
&d_R^2-B'^2-B'^1-\overline{Q}_L^2& \nonumber \\
&d_R^3-B''^1-\overline{Q}_L^3& \nonumber
\end{eqnarray}

\begin{table}[tbp]
\small
\begin{center}
\vspace{.5 cm}
\begin{tabular}{|c|c c c c|}
\hline
 $Q_{\fn}$ & -2    & -1     & 0       & 1      \\ 
\hline
up-type    & $T^1$ & $T^2$  & $T^3$   & $T^4$  \\  
           &       & $T'^1$ & $T'^2$  &        \\    
\hline
down-type  & $B^1$ & $B^2$  & $B^3$   &        \\  
           &       & $B'^1$ & $B'^2$  &        \\
           &       &        & $B''^1$ &        \\
\hline
\end{tabular}
\end{center}
\caption{$U(1)_{\fn}$ charges for VLQ's in model 2. Total $6+6=27$ heavy quarks are needed in this model. \label{table:table2}}
\end{table}

\noindent The $U(1)_{\fn}$ charge assignments are listed in table \ref{table:table2}. Clearly, model 2 has a simpler particle content than model 1, although this advantage comes at the expense of a larger scale separation between $v_s$ and $M$. 
The mass matrices are,
\begin{subequations}\label{subeq:model2}
\begin{equation}
\mathcal{M}^u = 
\begin{psmallmatrix}
0& 0& 0& \hu& 0& 0& 0& 0& 0\\
0& 0& 0& 0& \hu& \hu& 0& 0& 0\\
0& 0& \hu& 0& 0& 0& \hu& \hu& 0\\
0& 0& 0& M& \s& \s& 0& 0& 0\\
0& 0& \s& 0& M& 0& \s& \s& 0\\
0& 0& \s& 0& 0& M& \s& \s& 0\\
0& \s& 0& 0& 0& 0& M& 0& \s\\
0& \s& 0& 0& 0& 0& 0& M& \s\\
\s& 0& 0& 0& 0& 0& 0& 0& M\\
\end{psmallmatrix},
\end{equation}

\begin{equation}
\mathcal{M}^d = 
\begin{psmallmatrix}
0& 0& 0& \h& 0& 0& 0& 0& 0\\
0& 0& 0& 0& \h& \h& 0& 0& 0\\
0& 0& 0& 0& 0& 0& \h& \h& \h\\
0& 0& 0& M& \s& \s& 0& 0& 0\\
0& 0& 0& 0& M& 0& \s& \s& \s\\
0& 0& 0& 0& 0& M& \s& \s& \s\\
\s& \s& \s& 0& 0& 0& M& 0& 0\\
\s& \s& \s& 0& 0& 0& 0& M& 0\\
\s& \s& \s& 0& 0& 0& 0& 0& M\\
\end{psmallmatrix}
\end{equation}
\end{subequations}

\subsection{Couplings of the UV theory}
Each entry of the matrices given in \eqref{subeq:model1} and \eqref{subeq:model2} is accompanied by an $\mathcal{O}(1)$ coupling $\eta^{u/d}_{ij}$ of the UV theory, that has been suppressed. We take the bare mass $M$ to be same for all VLQ's, so that,  
\begin{equation}
\eta^{u/d}_{ii} = 1,~ \forall~ i>3.
\end{equation}

\noindent The rest of the yukawa couplings are related to the $\mathcal{O}(1)$ couplings $y^{u/d}_{ij}$ described in \eqref{eq: yij} by relations of the form,
\begin{equation}\label{eq: yuk_highlow}
y^{u/d}_{ij} \sim \prod\eta^{u/d}_{kl},
\end{equation}
where $\eta^{u/d}_{kl}$ are the relevant couplings appearing in the chain diagrams corresponding to $y^{u/d}_{ij}$. Although these couplings are $\mathcal{O}(1)$, their exact value is  unimportant for our purpose. The simplest possible choice is to keep all the yukawas unity. Another way is to choose them from a uniform distribution between say, $[1/3,3]$. Here we take all yukawas to be equal and parameterize them as,
\begin{equation} \label{eq:scaling}
\eta^{u/d}_{ij} = y,
\end{equation}
where $y$, chosen to be a real number, is a scaling factor, which we treat as a free parameter characterizing the overall yukawa strength. This simplifying assumption is reasonable, and the factor $y$ will be useful for book-keeping. In the limit $y=0$, the VLQ's completely decouple from the scalar sector, while the perturbativity of yukawas requires that $|y| \leq \sqrt{4\pi}$. We make a further simplification by choosing $y\geq 0$. 

\section{Effective potential}\label{sec: potential}
To study the flavon PT, we need to know the temperature-dependent effective potential, which is defined in terms of the generating functional of 1PI connected diagrams. For pedagogic reviews on the finite tmperature effective potential, see \citep{Carrington:1991hz,Quiros:1999jp,Laine:2016hma}. 

At zero temperature, the effective potential for a scalar field $\phi$ can be written in a loop expansion,
\begin{equation}
V(\phi)|_{T=0} = V_0(\phi) + V_1(\phi) + V_2(\phi) + \cdots ,
\end{equation} 
where the subscript indicates the loop order. The first term on the R.H.S., $V_0$ is the `classical' or tree-level contribution, while the first quantum corrections begin at the one-loop level. The temperature dependence on the other hand, is induced at one-loop level. Hence the one-loop temperature-dependent effective potential can be written schematically as:
\begin{equation}
V(\phi,T) = V_0(\phi) + V_1(\phi) + V_{1T}(\phi,T)
\end{equation}
Below, we compute each contribution to the effective potential separately.

\subsection{Tree-level potential}
The Higgs-flavon scalar potential is essentially an extension of the SM by a complex scalar (CxSM) \cite{Gonderinger:2012rd}, with a global $U(1)_{\fn}$ symmetry,
\begin{eqnarray}
    V_0(S,H) &=& -\mu_H^2 |H|^2  + \lambda_H |H|^4 + \lambda_{HS}|S|^2|H|^2 \nonumber\\
              &-& \mu_S^2 |S|^2 + \lambda_S|S|^4,
\end{eqnarray}

\noindent where all parameters are assumed to be real ($CP$ conserving potential). Tree-level vacuum stability requires,

\begin{equation}
\lambda_{H}\geq 0,~\lambda_{S} \geq 0.
\end{equation}

\noindent Further, we take $\lambda_{HS} \geq 0.$
\noindent Write the fields as: $$ H =  \begin{pmatrix}
G_{+}\\
\frac{h+i G_0}{\sqrt{2}}\\
\end{pmatrix}, ~~  S = \frac{1}{\sqrt{2}}s ~e^{i\rho},
$$ 
where $h,~s$ are real scalar fields, $G_0,~G_{\pm}$ are SM goldstone bosons, and $\rho$ is a pseudocalar. 

In recent years, it has become popular to associate the pseudoscalar $\rho$ with an Axion-like Particle (ALP), to additionally address the strong $CP$ problem of SM. The pseudoscalar field in this case has been variously called as the `flavorful axion' \cite{Wilczek:1982rv}, the `flaxion' \cite{Ema:2016ops} or the `axiflavon' \cite{Calibbi:2016hwq}. The ALP can acquire a tiny mass via the QCD anomaly, making it a pseudo-Nambu Goldstone (pNG) boson. Alternately, an explicit $U(1)_{\fn}$ breaking term such as $V_0(S,H)\supset \lambda_1(S^2+S^{*2})$ can generate a small mass term for $\rho$, with the mass scale constrained by low energy phenomenology \cite{huntingflavon}. The pseudoscalar plays no role in the dynamics of the flavon phase transition, which is governed by the real scalar field $s$, so we will not comment on it further.

When written in terms of field components, the potential also has a remnant  $Z_2$ symmetry where $s\rightarrow-s$. Spontaneous breaking of such a discrete symmetry by the VEV of $s$ can lead to formation of domain walls, which are disallowed by cosmology as they would eventually dominate the energy density of the universe, and directly contradict the cosmological principle. An additional $Z_2$-breaking term $V_0(S,H)\supset \lambda_2(S+S^*)$,
ensures that domain walls disintegrate fast enough to respect cosmological constraints. The contribution of these explicit symmetry breaking terms is however small and insignificant for analysing the nature of the flavon PT.

In terms of components, the potential is, 
\begin{eqnarray}
V_0(h,s,G_0,G_{\pm}) &=& -\frac{\mu^2_H}{2} \big(h^2+G_0^2+ 2G_+G_-\big) + \frac{\lambda_H}{4} \big(h^2+G_0^2+ 2G_+G_-\big)^2\nonumber\\
&+& \frac{\lambda_{HS}}{4}s^2\big(h^2+G_0^2+ 2G_+G_-\big)
- \frac{\mu^2_S}{2} s^2 + \frac{\lambda_S}{4} s^4.
\end{eqnarray}
Setting the goldstones to zero,
\begin{equation}
V_0(h,s) = -\frac{\mu^2_H}{2} h^2 + \frac{\lambda_H}{4} h^4 +\frac{\lambda_{HS}}{4}h^2s^2- \frac{\mu^2_S}{2} s^2 + \frac{\lambda_S}{4} s^4.
\end{equation}
Suppose at zero temperature the fields get a VEV: $$\langle h \rangle = v_h,~~\langle s \rangle = v_s.$$
We take $\mu^2_H$, $\lambda_{HS}$, $\lambda_S$ and $v_s$ as free parameters. 
 Minimization conditions are,
\begin{subequations}
\begin{equation}
\left.\frac{\partial V_0}{\partial h}\right\vert_{(v_h,v_s)} = 0,
\end{equation}
\begin{equation}
\left.\frac{\partial V_0}{\partial s}\right\vert_{(v_h,v_s)} = 0.
\end{equation}
\end{subequations}
These imply,
\begin{subequations}
\begin{equation}\label{min1}
\big(-\mu^2_H + \lambda_H v_h^2 +\frac{\lambda_{HS}}{2} v_s^2\big)v_h = 0,
\end{equation}
\begin{equation}
\big(-\mu^2_S + \lambda_S v_s^2 +\frac{\lambda_{HS}}{2}v_h^2\big)v_s = 0.
\end{equation}
\end{subequations}
Since $v_s\neq 0$, the second equation gives, 
\begin{eqnarray}\label{min2}
\mu_S^2 = \lambda_S v_s^2 + \frac{\lambda_{HS}}{2}v_h^2,
\end{eqnarray}
while in the first equation, we set $v_h = 0$, so that $\mu_H^2$ is an undetermined parameter. 
The scalar mass matrix is computed from the Hessian of the potential. Write $(h,s) = (\phi_1,\phi_2)$, then,
$$m^2_{ij} = \left.\frac{\partial^2 V_0}{\partial \phi_i\partial\phi_j}\right\vert_{(v_h,v_s)}$$
Setting $v_h=0$, 
\begin{gather}
\mathcal{M}^2_{\rm{scalar}} = \begin{pmatrix}
-\mu^2_H+\frac{\lambda_{HS}}{2} v_s^2 & 0 \\
0 & 2\lambda_S v_s^2 \\
\end{pmatrix}.
\end{gather}

As physical masses must be real, which gives us an additional constraint,
\begin{equation}
\mu^2_H\leq \frac{\lambda_{HS}}{2} v_s^2.
\end{equation}

We make the choice,
\begin{equation}
\mu^2_H = 0.01 ~\lambda_{HS} v_s^2,
\end{equation}
which numerically ensures that if a PT is possible, it occurs along the $s$ direction at one-loop level. This is however not the only possible choice, and it was verified that changing the multipliying coefficient upto an order of magnitude doesn't affect our results significantly. With the above choice, we can set $h=0$ in the potential in the rest of the analysis.
\begin{equation}
V_0(s) = -\frac{\mu_S^2}{2} ~s^2 + \frac{\lambda_S}{4}~s^4.
\end{equation}

The minimization condition \eqref{min2} now becomes,
\begin{equation}
\mu^2_S = \lambda_S v_s^2.
\end{equation}

\subsection{One-loop correction}
One-loop correction to the effective potential is given by the Coleman-Weinberg (CW) formula \cite{ColWein}. In the $\overline{\rm{MS}}$ scheme, in Landau gauge,
\begin{equation}\label{eq:ColWein}
V_{\rm{CW}}(h,s) = \frac{1}{64\pi^2}\sum_{i} (-1)^{f_i}n_i m^4_i(h,s)\bigg[\log\bigg(\frac{m^2_i(h,s)}{\mu^2}\bigg)- c_i\bigg],
\end{equation}
where the sum runs over all particles that couple to the flavon\footnote{We retain the field $h$ for the sake of completeness, although finally $h$ is to be set to zero.}. These include the SM Higgs ($h$), the SM  goldstones ($G_0,G_{\pm}$), the flavon ($s$), and the SM quarks and VLQ's ($F$). As the SM gauge bosons do not directly couple to $s$, they do not contribute to the one-loop effective potential. In the above equation, $\mu$ is the renormalization scale, which we set to $\mu^2 = v_s^2$. The factor $f_i$ is $+1$ $(0)$ for fermions (bosons). $n_i$ is the number of degrees of freedom: (1,1,3,12) for ($h,~s,~G_{0,{\pm}},~F$), and $c_i = 5/6$ for gauge bosons and $c_i = 3/2$ for others.

The field-dependent masses, $m_i^2(h,s)$, are obtained by expanding the tree-level Lagrangian around a constant field value $(h,s)$, and reading off the respective mass terms for the respective species. For the scalars, we get,
\begin{subequations}
\begin{equation}\label{eq: m2hs}
\mathcal{M}^2(h,s) = \begin{pmatrix}
-\mu^2_H + 3\lambda_H h^2 + \frac{\lambda_{HS}}{2}s^2 & \lambda_{HS} hs\\
\lambda_{HS} hs & \lambda_S (3s^2-v_s^2) + \frac{\lambda_{HS}}{2}h^2\\
\end{pmatrix},
\end{equation}
\begin{equation}\label{eq: m2gg}
m^2_{G_{0,\pm}}(s) = -\mu^2_H + \lambda_H h^2 + \frac{\lambda_{HS}}{2} s^2.
\end{equation}
\end{subequations}
Note that $m^2_i(s)<0$ for above scalars near the origin, implying an imaginary field-dependent mass. This well known problem occurs for scalars associated with a spontaneously broken symmetry, leading to a complex potential. The imaginary part of the potential is related to particle decay \cite{Weinberg:1987vp}. However, the dynamics of PT  is governed only by the real part of the potential and thus we always take the real part of the potential in our analysis.

\begin{figure}[tbp]
\centering % \begin{center}/\end{center} takes some additional vertical space
\includegraphics[width=.55\textwidth]{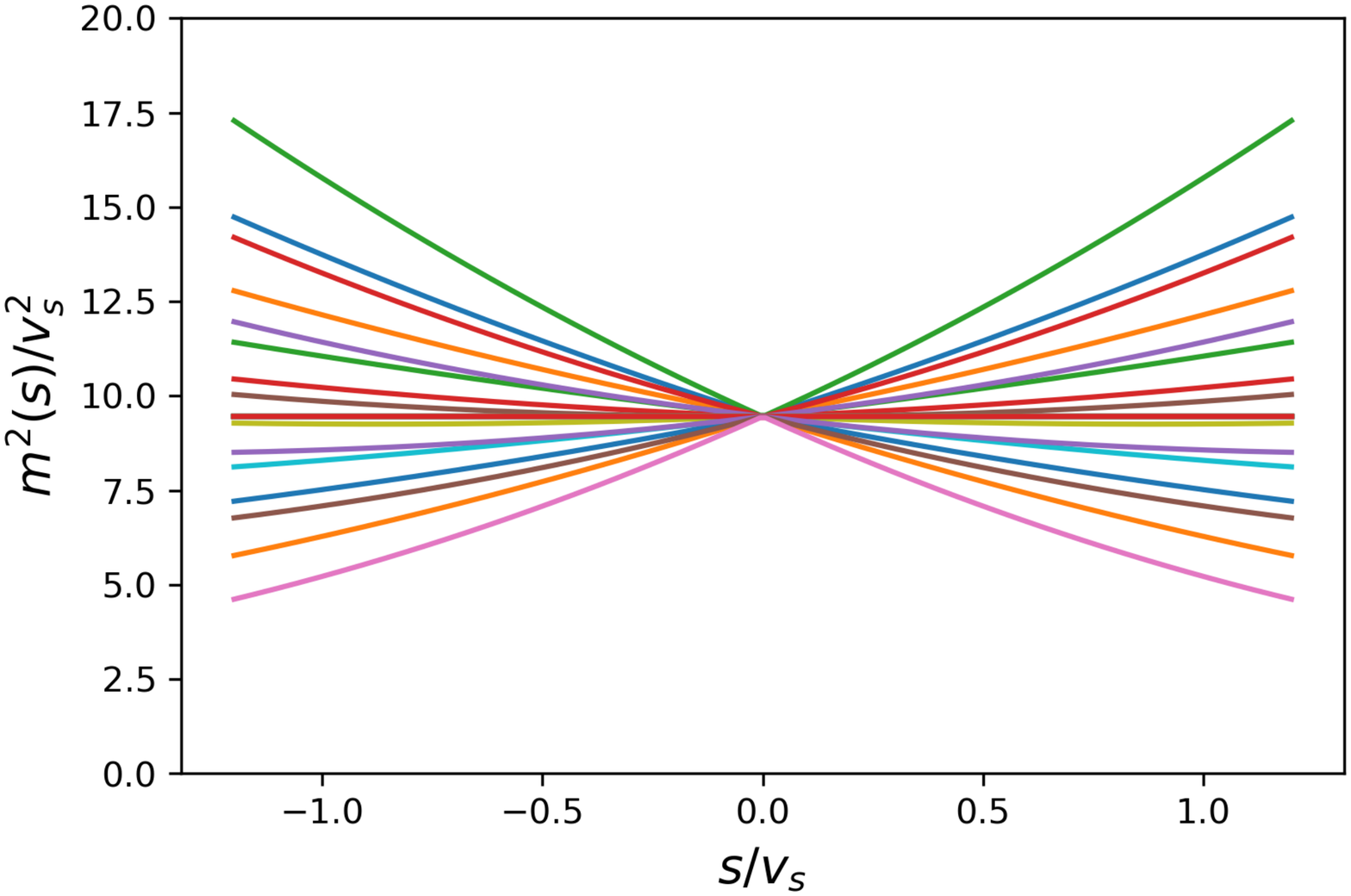}
\hfill
\caption{\label{fig: m_vlq} Field-dependent mass squared for the VLQ's of model 1. Here, we have taken $y=0.5$ and $v_s = 10^5$ GeV. The cyan coloured vertical line indicates $s=v_s$, where the physical VLQ masses are defined.}
\end{figure}

For fermions, the field-dependent mass matrices are obtained from the physical mass matrices $\mathcal{M}^{u/d}$ given in \eqref{subeq:model1} and \eqref{subeq:model2}, by the substitution:
$$ \hu\rightarrow \frac{1}{\sqrt{2}}\begin{pmatrix}
h\\
0\\
\end{pmatrix},~\h\rightarrow \frac{1}{\sqrt{2}}\begin{pmatrix}
0\\
h\\
\end{pmatrix},~\s\rightarrow \frac{s}{\sqrt{2}}.$$ 
$m_i^2(h,s)$ are then given by singular values of the field-dependent $\mathcal{M}^{u/d}$. For matrices of dimension larger than 5, the eigenvalues need to be evaluated numerically. The lightest three eigenvalues correspond to the SM quarks, while the rest are the VLQ's. Contribution of the light quarks to the potential is negligible and is therefore ignored. In figure \ref{fig: m_vlq}, we plot the squared field-dependent masses for the VLQ's of model 1. In the region $0\le s\lesssim M$, with $h=0$, the VLQ masses are well fitted by quadratic polynomials,
\begin{equation}\label{eq:VLQ param}
m^2_i(s) = M^2 + a_i~yMs + b_i~y^2s^2,
\end{equation}

where $a_i$ and $b_i$ are dimensionless fitting parameters, and $y$ is the scaling factor described in \eqref{eq:scaling}. For the largest and smallest VLQ masses, the parameters were found to be $|a_i|,~|b_i|\sim 1$, while they are smaller for others.

The $\overline{\rm{MS}}$ renormalization takes care of the UV divergent terms arising from integrals over momenta running in the loop.  However, additional finite counter-terms can be added to $V_{\rm{CW}}$ so that the VEV and flavon mass are unchanged from their tree-level values. Write,
\begin{subequations}
\begin{equation}\label{eq: ct1}
V_1(s) = V_{\cw}(s) + V_{\rm{c.t.}}(s),
\end{equation}
\begin{equation}\label{eq: ct2}
V_{\rm{c.t.}}(s) = -\frac{\delta\mu^2_s}{2} s^2 + \frac{\delta\lambda_s}{4} s^4.
\end{equation}
\end{subequations}
This is the so-called `On-shell renormalization' scheme. From \eqref{eq: ct1} and \eqref{eq: ct2}, the finite renormalization conditions are: 
\begin{subequations}
\begin{equation}
\left.\frac{\partial V_1}{\partial s}\right\vert_{v_s} = 0,
\end{equation}
\begin{equation}
\left.\frac{\partial^2 V_1}{\partial s^2}\right\vert_{v_s} = 0.
\end{equation}
\end{subequations}

These determine the parameters $\delta\mu^2_s$ and $\delta\lambda_s$,
\begin{subequations}
\begin{equation}
\delta\mu^2_s = \frac{1}{2 v_s}\left(3\left.\frac{\partial V_{\cw}}{\partial s}\right\vert_{v_s}-v_s \left.\frac{\partial^2 V_{\cw}}{\partial s^2}\right\vert_{v_s}\right),
\end{equation}
\begin{equation}
\delta\lambda_s = \frac{1}{2 v_s^3}\left(\left.\frac{\partial V_{\cw}}{\partial s}\right\vert_{v_s}-v_s\left.\frac{\partial^2 V_{\cw}}{\partial s^2}\right\vert_{v_s}\right).
\end{equation}
\end{subequations}

The scale-independence of the full effective potential \cite{ColWein} is spoiled by the appearance of $\mu$ in the CW formula. This is a result of truncating the potential at one-loop. Approximate scale-independence at one-loop is recovered when the Logs are $\mathcal{O}(1)$. Hence the renormalization scale needs to be chosen wisely, depending upon the regime of interest.

In the scenario considered here, there are two scales: $v_s$ and the VLQ mass scale $M$, with the scale separation quantified by the ratio,
\begin{equation}\label{eq: epsilon}
\epsilon \sim \frac{1}{\sqrt{2}}\frac{v_s}{M}.
\end{equation}
For model 1, $\epsilon = 0.23$, while for model 2, $\epsilon = 0.05$, implying a larger scale separation for model 2. To study the PT, we are interested in the behaviour of the potential around $s\sim v_s$. For model 1, the largest Log is $\sim 2.5$, while for model 2, the largest Log is $\sim 5.5$ at $\mu = v_s$. Hence the Logs are $\mathcal{O}(1)$. We treat these Logs as small, and analyse the one-loop potential at the fixed scale $\mu = v_s$, as the detailed RGE analysis is beyond the scope of this work.

\subsection{Finite temperature contribution}
 The one-loop finite temperature correction is essentially the free energy associated with the scalar field $\phi$ \cite{Carrington:1991hz}, 
\begin{equation}
V_{1T}(\phi,T) = \sum_i n_i\frac{T^4}{2\pi^2} J_{b/f}\bigg(\frac{m^2_i(\phi)}{T^2}\bigg),
\end{equation}
where the function $J_b$ ($J_f$) is defined for bosons (fermions). Write $x^2 \equiv \frac{m^2_i}{T^2}$, then $J_{b/f}$ is given by,
\begin{equation}
J_{b/f}(x^2) = \pm \int_0^{\infty} dy~ y^2 \log\big(1\mp e^{-\sqrt{x^2+y^2}}\big). 
\end{equation} 
These functions have well known  expansions for $x^2<<1$, and $x^2>>1$ . For small $x^2$ \cite{Cline_1997}, 
\begin{eqnarray}%\label{eq: highT}
J_f(x^2,n) = &-&\frac{7\pi^4}{360} + \frac{\pi^2}{24} x^2 + \frac{1}{32}x^4\big(\log x^2- c_f\big)\nonumber\\
&-&\pi^2x^2\sum_{l=2}^n\left(-\frac{1}{4\pi^2} x^2\right)^l\frac{(2l-3)!!\zeta(2l-1)}{(2l)!!(l+1)}\left(2^{2l-1}-1\right),\label{eq: highTf}\\
J_b(x^2,n) = &-&\frac{\pi^4}{45} + \frac{\pi^2}{12} x^2 - \frac{\pi}{6}\big(x^2\big)^{3/2}-\frac{1}{32}x^4\big(\log x^2- c_b\big)\nonumber\\
&+&\pi^2x^2\sum_{l=2}^n\left(-\frac{1}{4\pi^2} x^2\right)^l\frac{(2l-3)!!\zeta(2l-1)}{(2l)!!(l+1)},\label{eq: highTb}
\end{eqnarray}
where, $c_+=3/2+2\log\pi-2\gamma_E$, and $c_-=c_++2\log 4$. Here $\gamma_E$ is the Euler-Mascheroni constant, and $\zeta(x)$ is the Riemann-$\zeta$ function. $n$ is the order till we want to expand the sum. Usually taking $n=3,4$ gives good results.

For large $x^2$, both fermions and bosons have the same expansion at order $n$ \cite{Cline_1997},
\begin{align}\label{eq:lowT}
J_{b/f}(x^2,n) = &-\exp\bigg(-(x^2)^{1/2}\bigg)\bigg(\frac{\pi}{2}(x^2)^{3/2}\bigg)^{1/2}\sum_{l=0}^n \frac{1}{2^ll!}\frac{\Gamma(5/2+l)}{\Gamma(5/2-l)}(x^2)^{-l/2},
\end{align}
where $\Gamma(x)$ is the Euler Gamma function. For intermediate values of $x^2$, the functions $J_{b/f}(x^2)$ can be obtained by interpolating between the large $x^2$ and small $x^2$ approximation. The finite temperature effects have been computed in this work using the CosmoTransitions package \cite{Wainwright:2011kj}.

\subsection{Daisy resummation}
The perturbative loop expansion of the effective potential breaks down near the critical temperature $T_c$ due to infrared divergences from the bosonic Matsubara zero-modes \cite{Carrington:1991hz}. Leading contribution to the divergences comes from multiloop `daisy diagrams', which needs to be resummed. We achieve this by making the substitution: $m^2_i(s)\rightarrow m^2_i(s) + \Pi_i(T)$ \cite{Parwani:1991gq}, for all the bosons, where $\Pi_i(T)$ is the squared thermal mass for species $i$. Thermal masses are obtained by substituting \eqref{eq: m2hs}, \eqref{eq: m2gg} in the high-$T$ expansion \eqref{eq: highTb} to leading order in $T$. These are:
\begin{eqnarray}\label{eq: daisy}
\Pi_{h,G_{0,\pm}}(T) &=& T^2\bigg(\frac{g_1^2}{16}+\frac{3 g_2^2}{16}+\frac{\lambda_{HS}}{12} + \frac{\lambda_H}{2} + \frac{y_t^2}{4}\bigg),\\
\Pi_{s}(T) &=& T^2\bigg(\frac{\lambda_{HS}}{6} + \frac{\lambda_S}{3}\bigg),
\end{eqnarray}
where $g_1,~g_2$ are the usual SM gauge couplings, and $y_t$ is the top yukawa coupling. All SM couplings are taken at the scale $\mu = v_s$, and are found by evolving the SM RGE's \cite{Buttazzo:2013uya}. 

Now that all the contributions to the effective potential are defined, we finally have:
\begin{equation}
V_{\rm{eff}}(s,T) = V_0(s) + V_{\cw}(s,T) + V_{{\rm{c.t.}}}(s) + V_{1T}(s,T),
\end{equation}
with the Daisy-corrected field-dependent masses.
\section{Vacuum stability}\label{sec: stability}
Fermions contribute with a negative sign to the CW formula \eqref{eq:ColWein}, and tend to destabilize the potential at large field values. In the context of FN, vacuum stability has been addressed earlier \citep{Giese:2019khs}. Similarly, Higgs vacuum stability of the SM in presence of vector-like fermions has also been discussed \citep{Blum:2015rpa,Gopalakrishna:2018uxn,Borah:2020nsz,Higaki:2019ojq}.  Typically, UV completions of FN contain several ($\mathcal{O}(10)$) heavy fermions with $\mathcal{O}(1)$ yukawa couplings to the flavon, destabilizing the effective potential. Here we discuss the vacuum stability along the flavon direction. 

The negative contribution of VLQ's can be seen from the effective quartic coupling at one-loop level. Setting the renormalization scale $\mu^2 = v_s^2$ and using \eqref{eq: ct1},
\begin{eqnarray}\label{eq: instability}
\frac{\lambda_{s,\rm{eff}}}{4} &=& \frac{1}{4!} \left.\frac{\partial^4}{\partial s^4}\big(V_0+V_1\big)\right\vert_{s=0}\nonumber\\
&\approx& \frac{\lambda_s+\delta\lambda_s}{4} + \frac{9}{64\pi^2}\log\big(\lambda_s\big)+\frac{\lambda^2_{HS}}{64\pi^2}\log\left(\frac{\lambda_{HS}}{100}\right)\nonumber\\
&+& \sum_{i\in {\rm{VLQ}}} \frac{y^4}{64\pi^2}
\bigg\{a_i^4-12 a^2_ib_i - 12b_i^2 \log \bigg(\frac{M^2}{v_s^2}\bigg)\bigg\},
\end{eqnarray}
where the second term on the RHS is the flavon contribution, the third term is contributed by $h,~G_{0,\pm}$, and we have used the fitting polynomials \eqref{eq:VLQ param} in the last term for the VLQ contribution. Since the coefficients $a_i$ and $b_i$ are all $\mathcal{O}(1)$, the VLQ term is dominated by the factors of 12, and hence the fermionic contribution to the quartic coupling is negative. Assuming for simplicity that each term in the VLQ sum contributes roughly equally, the total contribution also increases with the number of VLQ's. Lastly, the VLQ contribution scales as $y^4$, therefore, increasing the yukawa factor by a small amount dramatically increases the negative contribution.

The instability scale, $\Lambda_{\rm{inst}}$ is estimated from the field value at which the potential falls below the height of the local minimum at $s=v_s$. 
For $\lambda_S\sim\mathcal{O}(0.1)$, $\lambda_{HS}\sim\mathcal{O}(1)$ and $y=1$, $\Lambda_{\rm{inst}}$ was found around the VLQ mass scale $M$. Increasing the yukawa scaling factor further pushes the instability scale lower. Naively, the potential can be stabilized by increasing the couplings ($\lambda_S$, $\lambda_{HS}$), to counteract the fermionic contribution. However, this procedure is limited by the perturbativity bounds on the quartic couplings: $\lambda_S, \lambda_{HS}\leq 4\pi$. For large couplings, the one-loop results also become unreliable, and any FOPT observed from the one-loop analysis can't be trusted. As is evident from \eqref{eq: instability}, the fermion contribution to the quartic coupling is highly sensitive to the scaling factor $y$. Hence we can push $\Lambda_{\rm{inst}}$ to at least a few times of $M$ by decreasing $y$. We found that by keeping $y = 0.5$, $\Lambda_{\rm{inst}}$ is atleast $~3M$ for the smallest value of the couplings $(\lambda_S,\lambda_{HS})$  in the parameter region of interest \footnote{Keeping $y=0.5$ for all couplings uniformly can contribute slightly to the hierarchy of fermion masses as some combinations $y_{ij}$ may be smaller than $\mathcal{O}(1)$ according to \eqref{eq: yuk_highlow}, but this effect is tolerable.}. Beyond the instability scale, the  potential can be stabilized by adding new heavy bosonic degrees of freedom which couple to the flavon as illustrated in Appendix \ref{app: bosons}. If the mass scale of the new bosons is high compared to $M$, we can write down an Effective Field Theory (EFT), by introducing higher dimensional operators in the flavon potential in a model-independent way (see for example \cite{Blum:2015rpa}),
\begin{equation}
V_{1} \rightarrow V_{1} + \frac{1}{\Lambda_b^2} |S|^6,
\end{equation}
where $\Lambda_b\sim \Lambda_{\rm{inst}}$ is the cut-off scale associated with the new bosons, and the Wilson coefficient is taken to be unity. For sufficiently heavy bosons, the thermal contribution is suppressed by the Boltzmann factor, as can be seen from the small $T$ expansion \eqref{eq:lowT}. Hence the heavy bosonic fields play an insignificant role in determining the nature and strength of the PT. We assume that vacuum stability at field values beyond $\Lambda_{\rm{inst}}$ is taken care of by such bosons, without bothering about any particular construction.

\section{First order flavon phase transition}
\label{sec:FOPT}

\subsection{Characterizing FOPT}
A FOPT proceeds via bubble nucleation, and is characterized by three parameters related to the latent heat, rate of PT, and the nucleation temperature, which we review below. As the universe cools and the unbroken vacuum becomes metastable, tiny  bubbles of the stable phase nucleate. In what follows, we take $T$ as the temperature of the universe at time $t$ after the Big Bang. The temperature at which the two minima become degenerate is called the critical temperature $T_c$, below which the nucleation rate per unit volume, $\Gamma(T)$, is given by \citep{Linde:1980tt,Kobakhidze_2017},
\begin{equation}\label{eq: tunnel_prob}
\Gamma(T) =\left(\frac{S_3(T)}{2\pi T}\right)^{3/2} T^4 e^{-S_3(T)/T},
\end{equation}
where $S_3(T)$ is the $O(3)$ symmetric Euclidean bounce action, for transitions at finite temperature. We calculate $S_3$ using the Python based CosmoTransitions \cite{Wainwright:2011kj} package. The tunnelling action $S_3$ decreases with time as the universe cools below $T=T_c$, thereby increasing the tunnelling probability. The nucleation temperature is defined as the temperature $T_n$, at which the tunneling probability per unit volume becomes  $\mathcal{O}(1)$, i.e., at $T=T_n$, at least one bubble gets nucleated per Hubble volume. This happens when, 
\begin{equation}
\Gamma(T_n)\approx \big(H(T_n)\big)^4.
\end{equation}
In the radiation dominated era, using \eqref{eq: tunnel_prob}, this implies the condition \cite{Huang:2020bbe},
\begin{equation}\label{eq: nuclCriterion}
\frac{S_3(T)}{T} \simeq -4 \ln\left(\frac{T}{m_{\rm{Pl}}}\right),
\end{equation}
where $m_{\rm{Pl}}$ is the Planck mass. 

The strength of FOPT is characterized by the parameter $\alpha$, given by the ratio of the vacuum energy density and the radiation density of the surrounding plasma. Let $T_*$ be the temperature around which GW production is most significant. Then,
\begin{equation}
\alpha = \left.\frac{\rho_{\rm{vac}}}{\rho_{\rm{rad}}}\right\vert_{T_*},
\end{equation} 
where $\rho_{\rm{vac}}$ is the energy density difference between the metastable and the stable phases at $T=T_*$: $\rho_{\rm{vac}}=\Delta \bigg(V-\frac{\partial V}{\partial T}\bigg)$. $\rho_{\rm{rad}}$ is the radiation density of the thermal bath given by,
\begin{equation}
\rho_{\rm{rad}}(T) = \frac{\pi^2}{30}g_{*}T^4.
\end{equation}
We choose $T_*\approx T_n$ in our calculations, since this is where significant  GW production is expected. Around $T_*$, $g_{*} = 107.75$, with $106.75$ contributed by the SM fields, and 1 contributed by the pseudoscalar $\rho$, while the rest of the fields remain non-relativistic.

Finally we need the parameter $\beta$, which characterizes the rate at which the PT proceeds.

\begin{equation}
\beta \equiv  -\left.\frac{dS}{dt}\right\vert_{t=t_*} = TH_*\left.\frac{dS}{dT}\right\vert_{T=T_*},
\end{equation}
where, $S=S_3/T$, and $H_*$ is the Hubble's constant at $T=T*$. The negative sign in the first equality is due to the fact that the action decreases with time.

\subsection{A case for flavon FOPT}
To generate an FOPT, a barrier must be induced between the two degenerate minima at $T_c$. Although the conventional wisdom is that only bosons contribute to the barrier to generate a SFOPT, it has been shown in the context of electroweak baryogenesis, that a SFOPT is also possible in purely fermionic extentions of the SM \citep{Carena_2005,Fairbairn_2013,
Davoudiasl_2013,Egana_Ugrinovic_2017,Angelescu_2019,Cao:2021yau}. It is therefore interesting to see the effect of the heavy VLQ's on the strength of FOPT in a flavon PT. Let us parametrize the finite temperature effective potential, $V_{\rm{eff}}$ as,
\begin{equation}
V_{\rm{eff}}(s,T) = \frac{\mu^2(T)}{2} s^2 + \frac{\lambda_3(T)}{3} s^3 + \frac{\lambda_4(T)}{4} s^4 + \frac{\lambda_6(T)}{6} s^6.
\end{equation}  
The following two cases may arise:
\begin{itemize}
\item \textbf{Case 1}: When the dimension 6 term is negligible, and 
$$\mu^2(T_c)>0,~\lambda_3(T_c)<0,~\lambda_4(T_c)>0.$$
The cubic term is primarily responsible for inducing the barrier. Such a cubic term is generated for example by bosons in the high $T$ expansion as in \eqref{eq: highTb}, while fermions do not induce any such contribution. However, both fermions and bosons have the same expansion in the low $T$ limit \eqref{eq:lowT}, and hence can contribute equally to the formation of a barrier \cite{Egana_Ugrinovic_2017}.

\item \textbf{Case 2}: If the cubic term is small compared to the others, then a barrier can also be formed when \cite{Grojean:2004xa},
$$\mu^2(T_c)>0,~\lambda_4(T_c)<0,~\lambda_6(T_c)>0.$$
Although the $|S|^6$ operator may be generated by integrating out the stabilizing heavy bosonic fields, we checked that it plays a negligible role in forming a barrier in our parameter space of interest.
\end{itemize}

\begin{figure}[tbp]
\centering % \begin{center}/\end{center} takes some additional vertical space
\includegraphics[width=.55\textwidth]{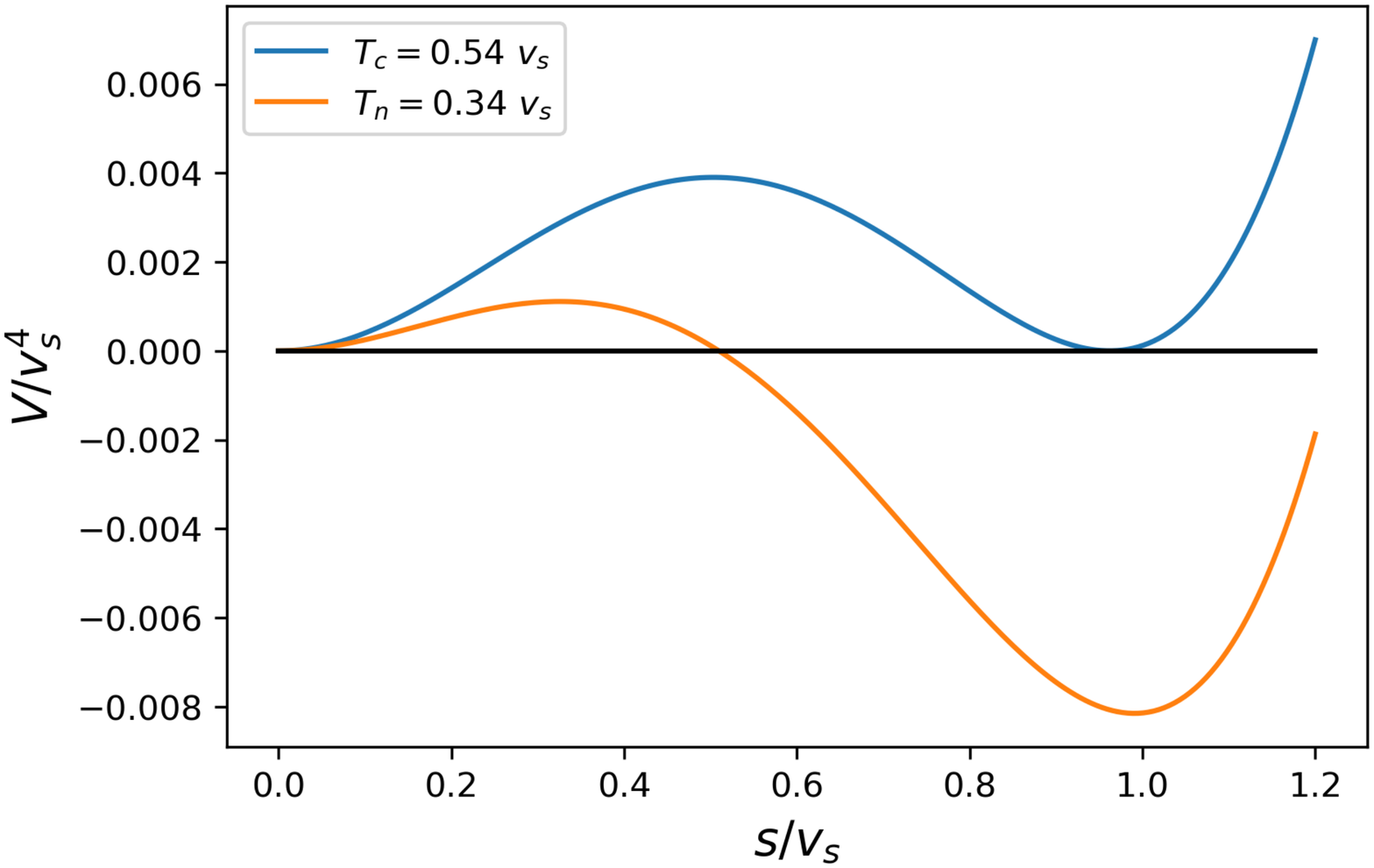}
\hfill
\caption{\label{fig: nearPT} Shape of the potential of model 1 near the FOPT, for $v_s=10^5$ GeV, $y=0.5$, $\lambda_S = 0.105$, $\lambda_{HS} = 4.253$.}
\end{figure}

Clearly the first case is relevant to our scenario, where the effect of a possible $|S|^6$ operator is taken to be small, as discussed in the previous section. In figure \ref{fig: nearPT}, we depict the barrier formed at $T=T_c$ (blue), and at $T=T_n$ (orange), for a benchmark point in model 1. The ratio, $\frac{v_s(T_n)}{T_n}\gtrsim 2.7$, indicates a SFOPT. At $T=T_n$, the bosons $\{h,s,G_{0,\pm}\}$ have masses $\lesssim T_n$, while the VLQ masses are larger than $T_n$.  As a result, the bosonic contribution to the effective potential is governed by the high-$T$ expansion \eqref{eq: highTb}, which incidently justifies the need for taking Daisy resummation into account. On the other hand, the fermionic contribution is governed by the low-$T$ expansion \eqref{eq:lowT}. In fact, this behaviour is observed in the entire region of parameter space considered in the next section. Therefore, barrier formation happens primarily by bosons via the cubic term in the high-$T$ expansion, and the fermionic contribution is comparatively small due to the Boltzmann suppression factor in \eqref{eq:lowT}. In the analysis below, the VLQ's are observed reduce the strength of FOPT slightly when compared to the no-VLQ case.

\subsection{Parameter space}
Now we discuss the prospects of achieving a SFOPT in the two FN models \eqref{eq: model1} and \eqref{eq: model2}. The parameters considered are, $\{y,v_s,\lambda_S,\lambda_{HS}\}$. The VLQ mass scale $M$ is directly related to $v_s$ via \eqref{eq: epsilon}, and is therefore not an independent parameter. 

In what follows, we keep $v_s = 10^5$ GeV, $y\in\{0,0.5\}$, and perform a parameter scan in the $\lambda_S-\lambda_{HS}$ plane, with $\lambda_S\in [0, 0.3]$, and $\lambda_{HS} \in [2.5,6]$. By restricting ourselves to this particular region of the plane, we ensure that perturbativity is obeyed, and that the one-loop calculations are reliable. For the two models, we choose $y=0.5$, along with the described lower bounds on $\lambda_S$ and $\lambda_{HS}$ so that the instability scale $\lambda_{\rm{inst}}$ is at least $\gtrsim 3M$. For each point on the plane, we explore the phase structure of the finite temperature potential using the `findallTransitions' and `calcTctrans' methods of the Python based package CosmoTransitions \cite{Wainwright:2011kj}. In case of an FOPT, the tunnelling action $S_3$ is computed using the class `pathDeformation', and $T_n$ is numerically obtained using the condition \eqref{eq: nuclCriterion}. Next, the other two PT parameters, namely $\alpha$ and $\beta/H_*$ are computed. To compute $\beta$, we plot the action around $T=T_n$, and calculate the first derivative using the fourth order finite difference formula. 

\begin{figure}[tbp]
\centering % \begin{center}/\end{center} takes some additional vertical space
\includegraphics[width=.32\textwidth]{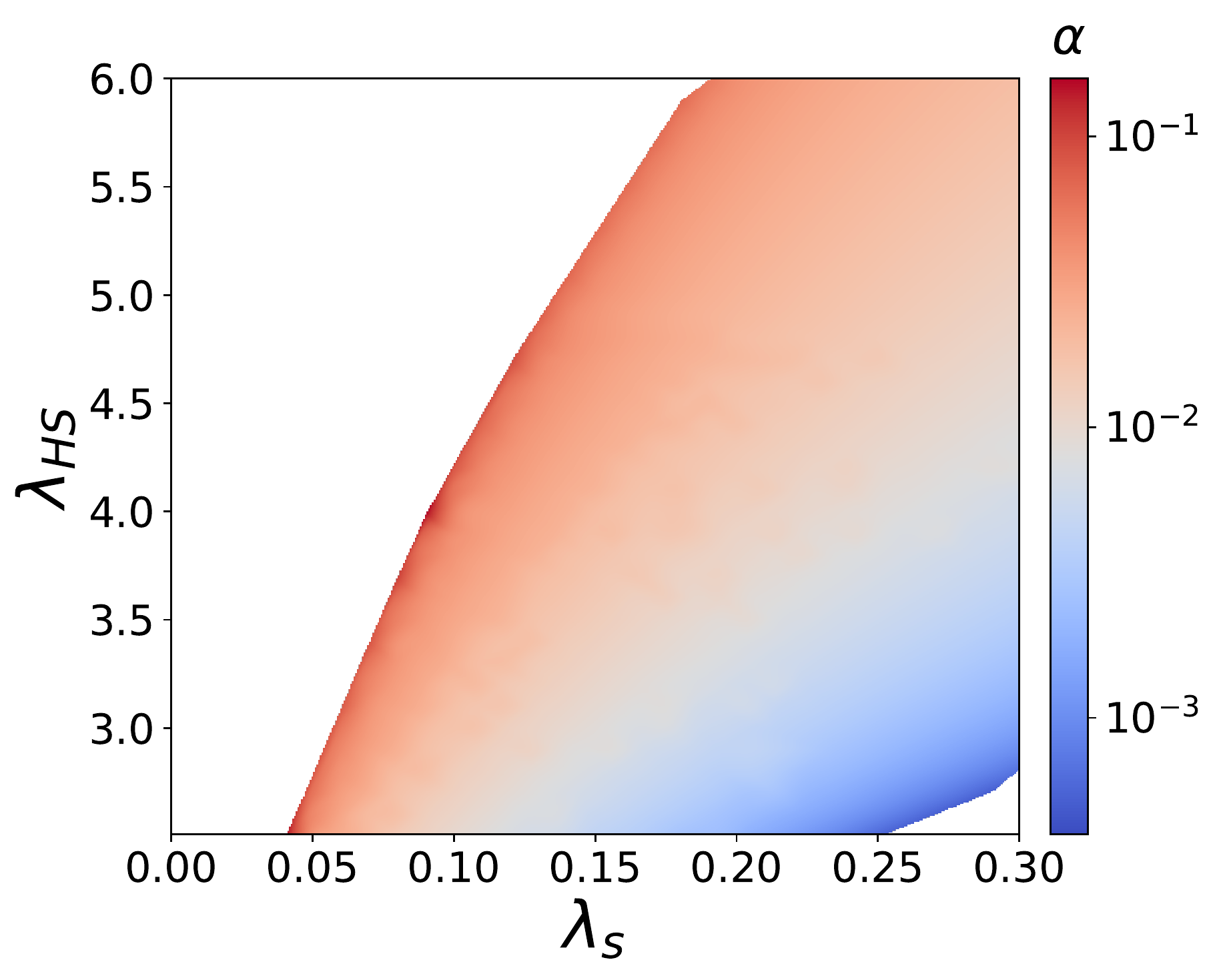}
\hfill
\includegraphics[width=.32\textwidth]{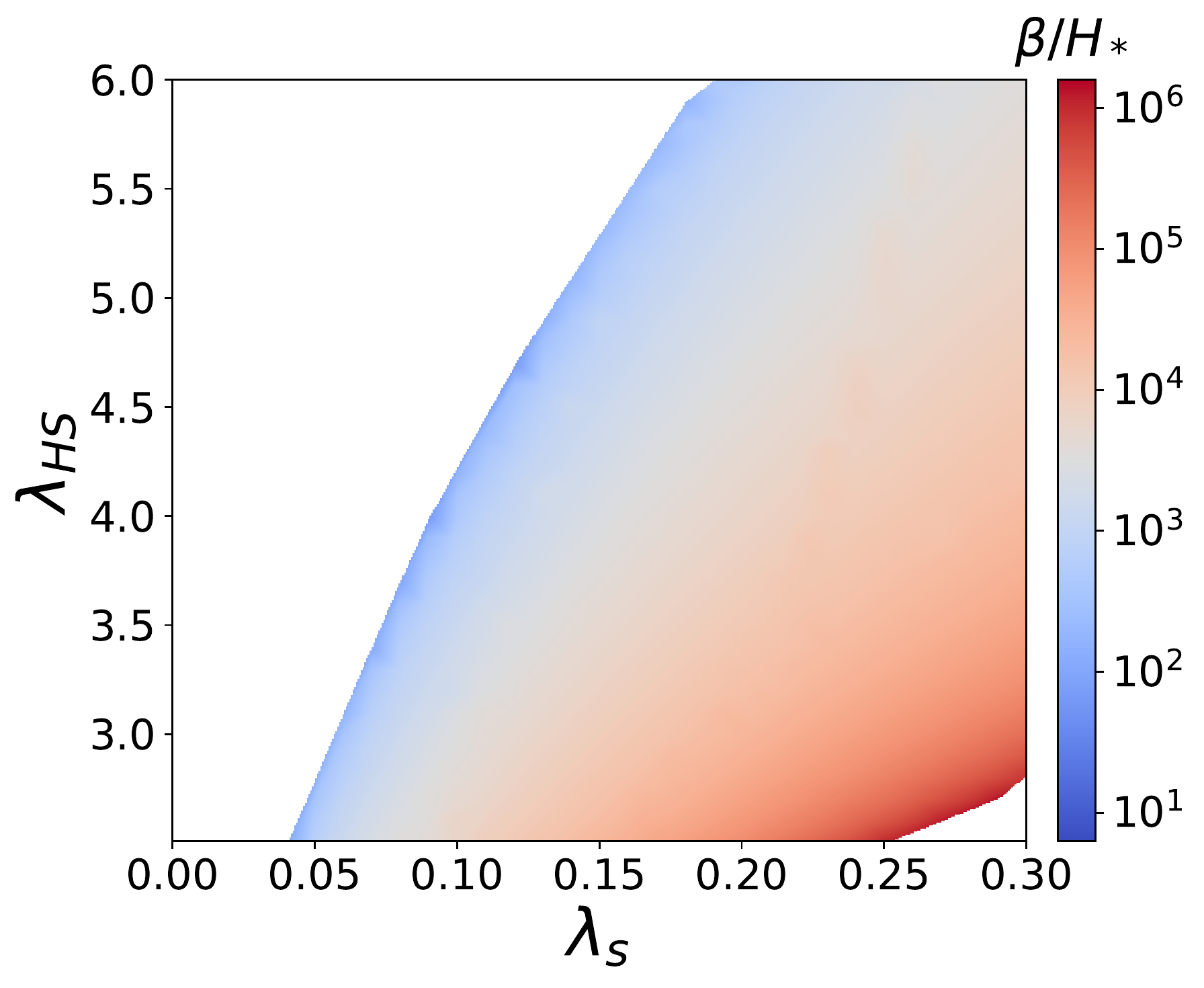}
\hfill
\includegraphics[width=.32\textwidth]{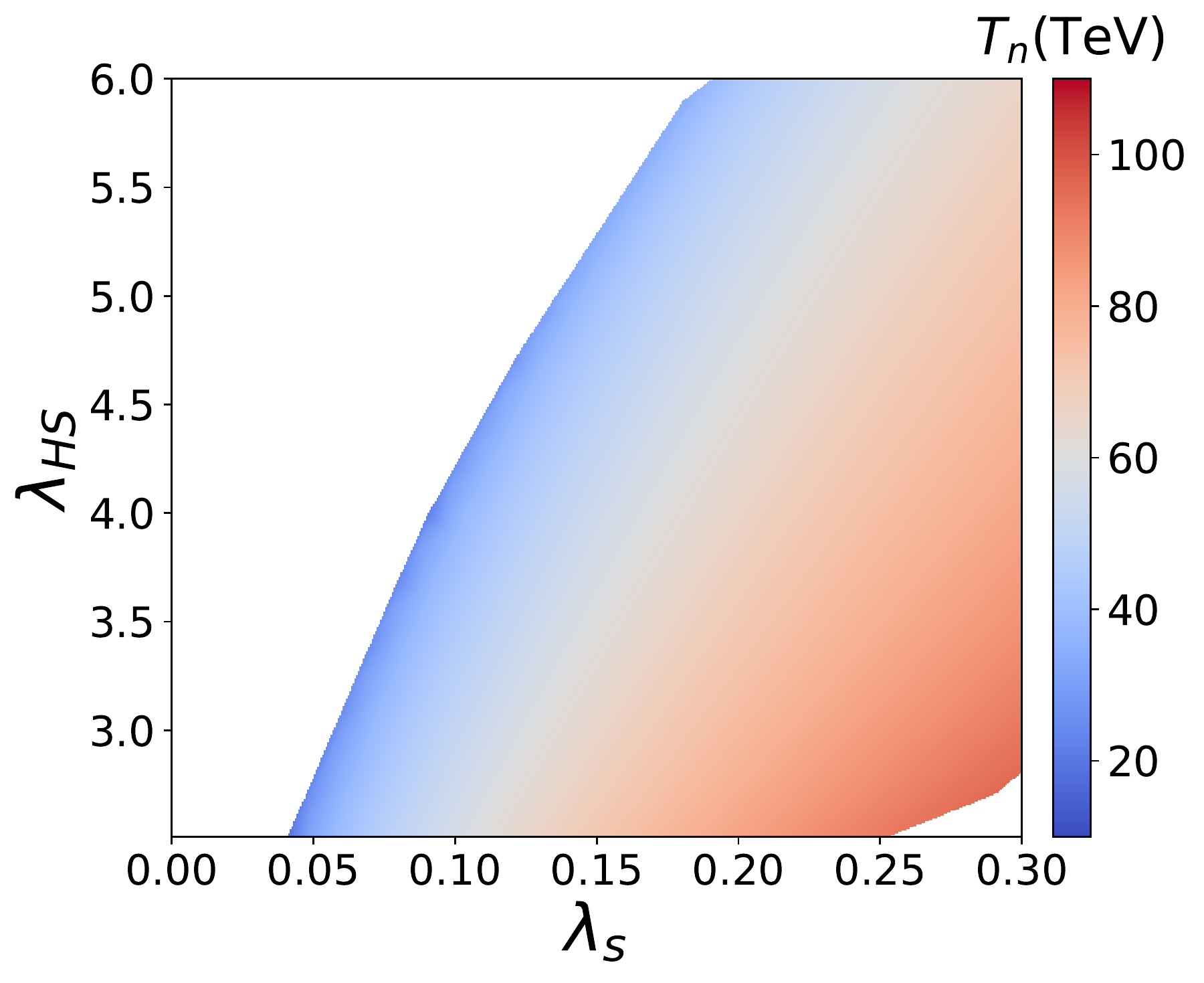}

\caption{\label{fig: noVLQ} FOPT parameter space when the influence of VLQ's is switched off. Colors show PT parameters: $\alpha$ (left), $\beta/H_*$ (middle), $T_n$ (right). Here, $y=0$, and $v_s = 10^5$ GeV.}
\end{figure}

First, let us examine the region of FOPT without taking the VLQ's into consideration, which corresponds to setting $y=0$ in either model 1 or model 2. Let us call it the `reference model'. The region of FOPT is shown in  figure \ref{fig: noVLQ}, where $v_s$ is set to $10^5$ GeV, as a function of $\alpha$ (left panel), $\beta/H_*$ (middle panel), and $T_n$ (right panel). The observed range of the PT parameters is:  $\alpha\in [10^{-3},10^{-1}]$, $\beta/H_* \in [10^1,10^6]$, and $T_n \in [10,100]$ TeV. A SFOPT occurs when $\alpha$ is large, and $\beta/H_*$ is small. The region of FOPT is defined by a sharp boundary on the left hand side of the $\lambda_S-\lambda_{HS}$ plane. In the white region beyond the left boundary, the metastable minimum at $s=0$, is deep enough, so that the tunnelling to the $s=v_s$ minimum never achieves completion, as the condition \eqref{eq: nuclCriterion} is never met. For a given $\lambda_{HS}$, the boundary occurs at the lowest value of $\lambda_S$ where the FOPT can reach completion, and this is where it is the strongest. As we move to the right along the $\lambda_S$-axis, the FOPT becomes progressively weaker, as shown by the plots in figure \ref{fig: noVLQ}. Beyond the boundary on the right hand side, the FOPT is either extremely weak, or the PT is second order. Notice that the nucleation temperature $T_n$ increases with increasing couplings $\lambda_S$ and $\lambda_{HS}$.

\begin{figure}[tbp]
\centering % \begin{center}/\end{center} takes some additional vertical space
\includegraphics[width=.32\textwidth]{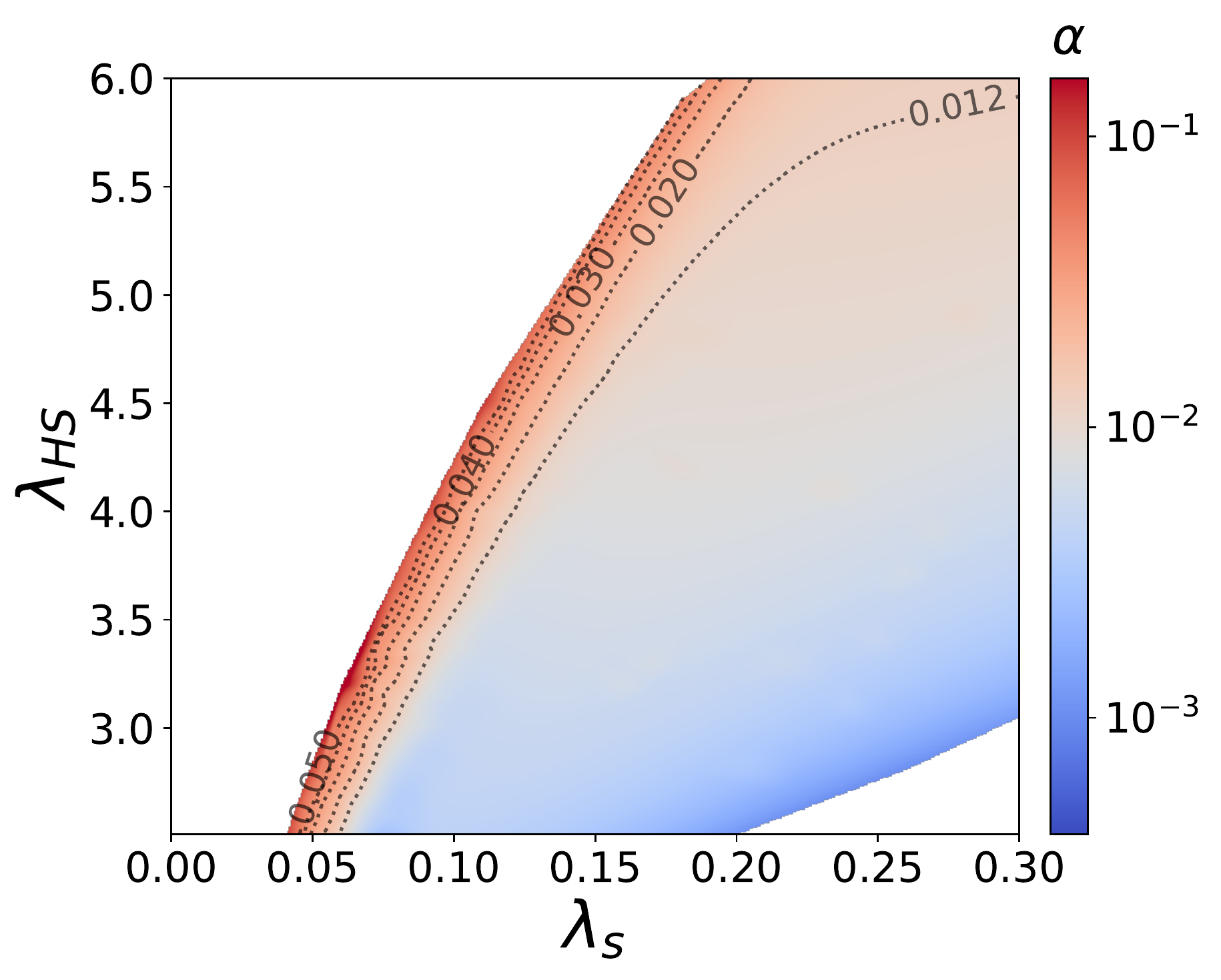}
\hfill
\includegraphics[width=.32\textwidth]{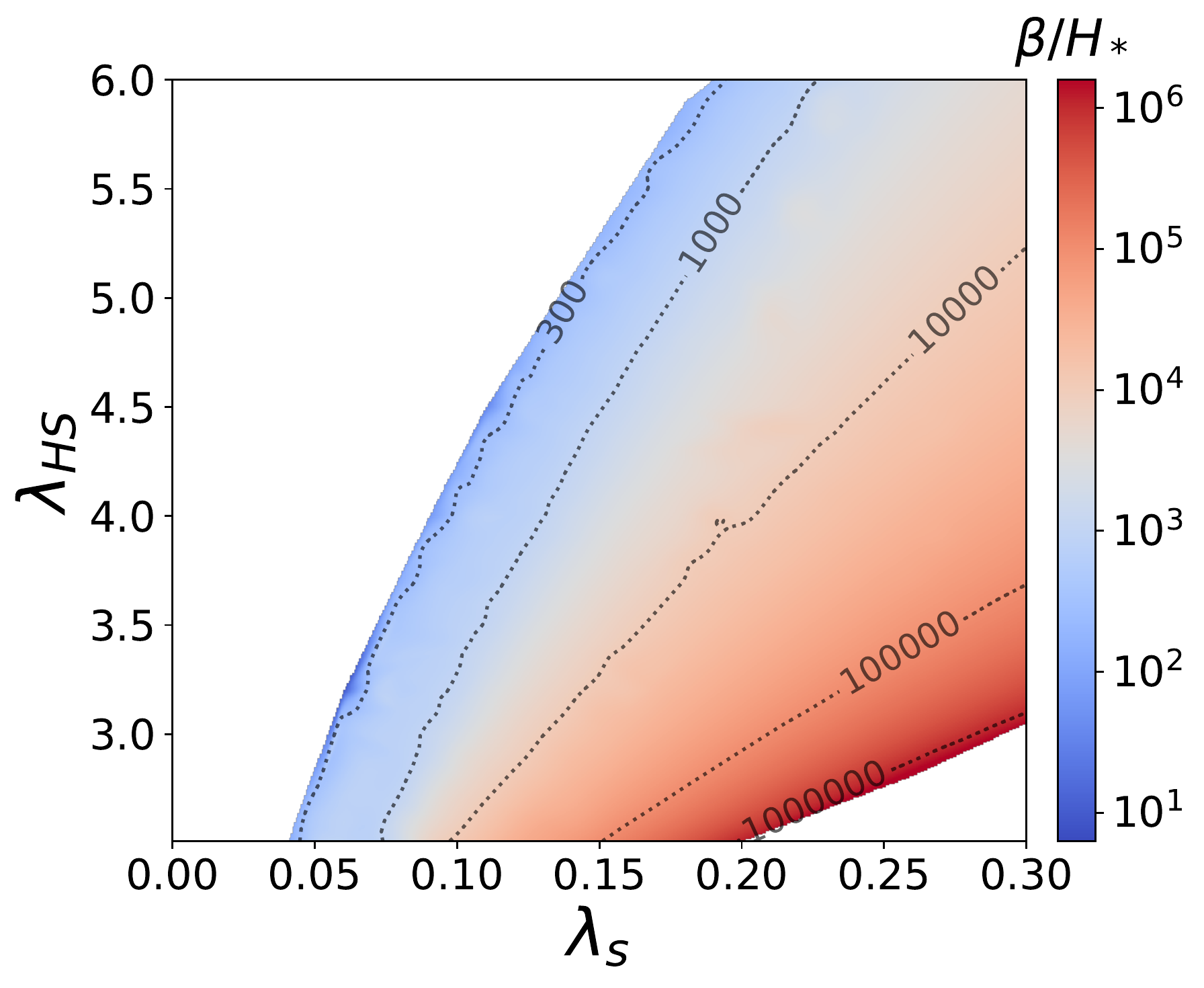}
\hfill
\includegraphics[width=.32\textwidth]{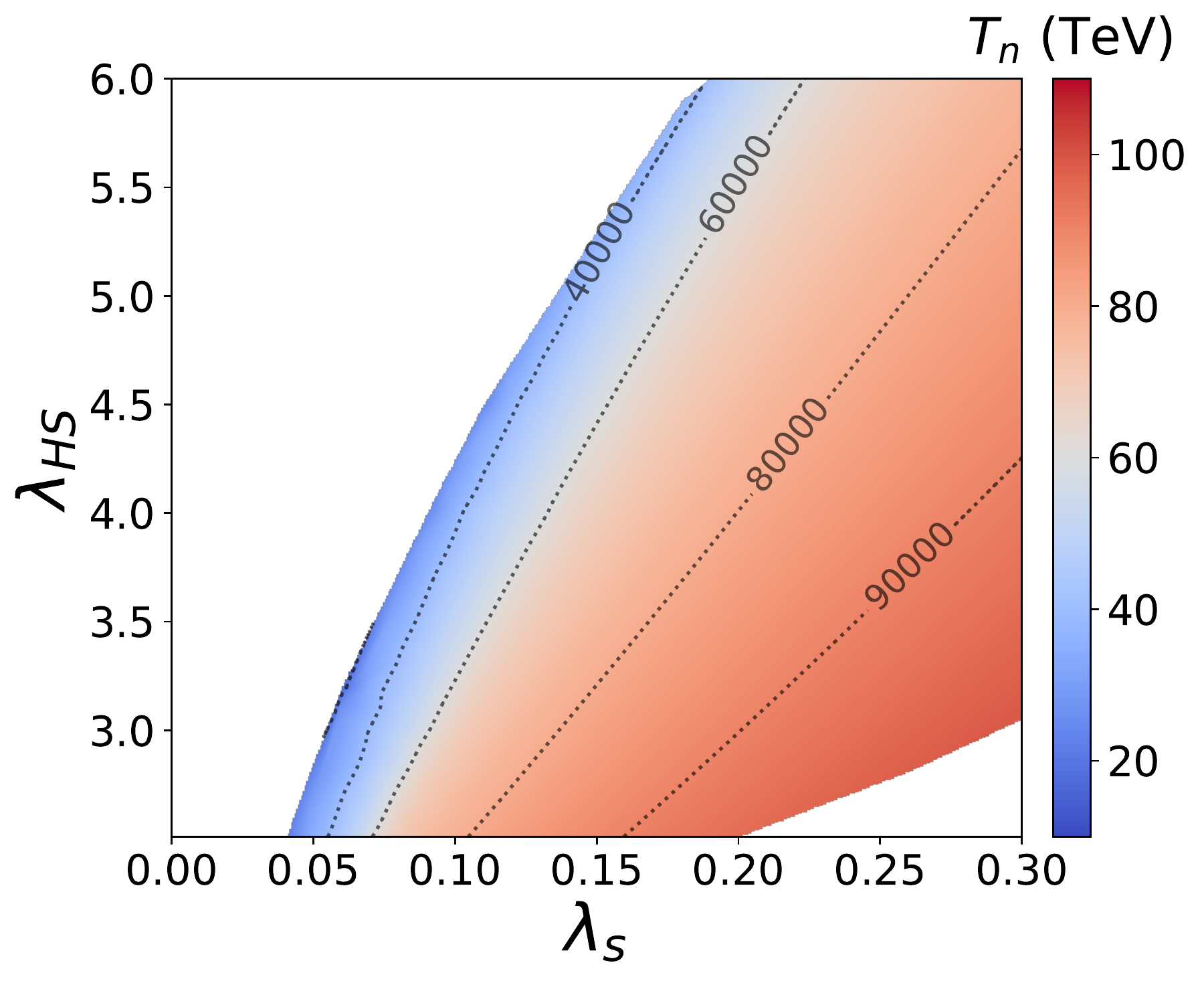}

\caption{\label{fig: param_scan1} PT parameters for model 1 with flavon VEV, $v_s=10^5$ GeV, and $y=0.5$. Left panel shows the variation of $\alpha$, with contours $\alpha=0.012,~0.02,~0.03,~0.04,~0.05$. In the middle panel, $\beta/H_*$ is depicted with contours $\beta/H_*= 3\times10^2,~ 10^3,10^4,~10^5,~10^6$. In the right panel, $T_n$ is depicted with contours $T_n=4,~6,~8,~9$ $\times 10^4$ GeV.}
\end{figure}

\begin{figure}[tbp]
\centering % \begin{center}/\end{center} takes some additional vertical space
\includegraphics[width=.32\textwidth]{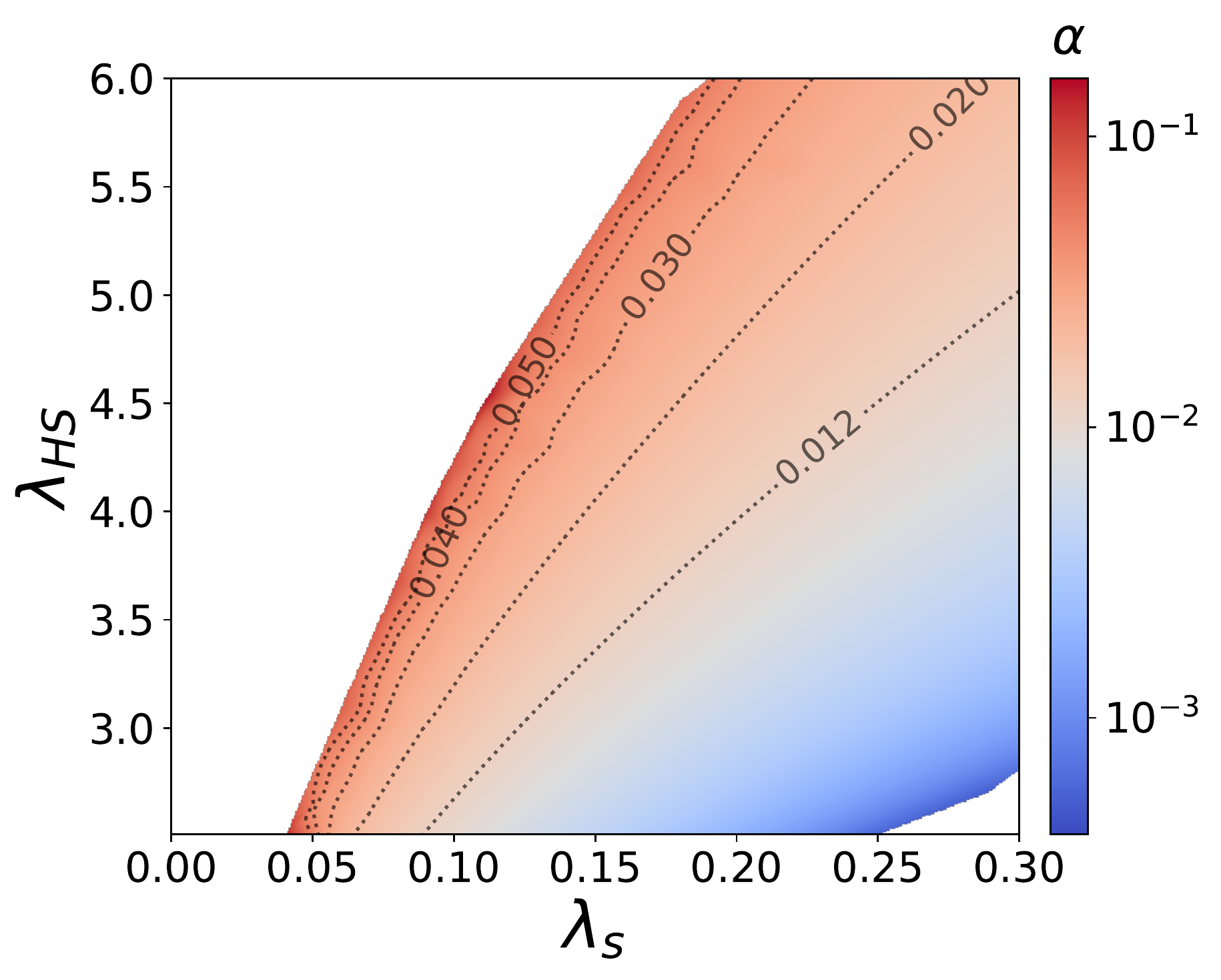}
\hfill
\includegraphics[width=.32\textwidth]{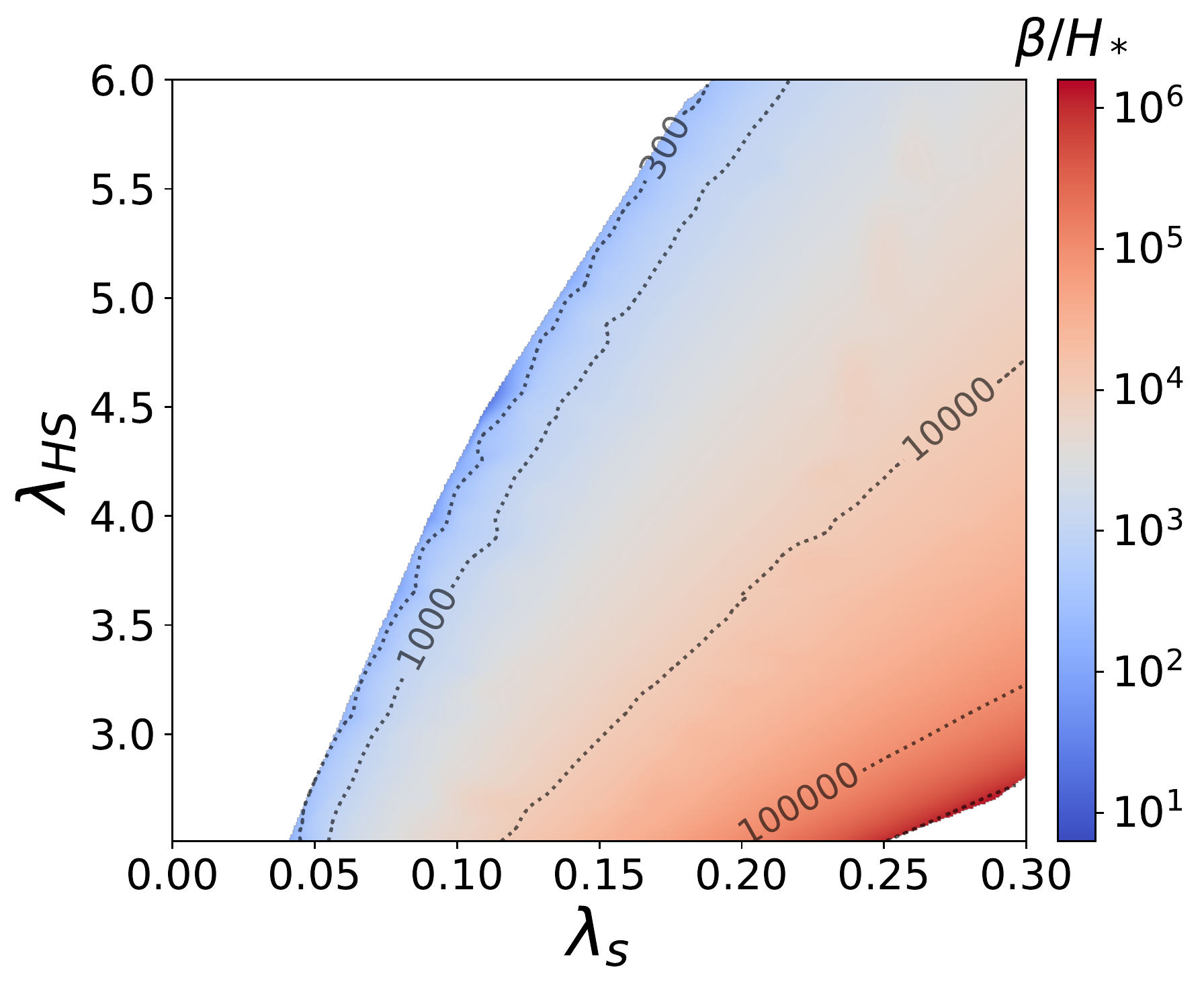}
\hfill
\includegraphics[width=.32\textwidth]{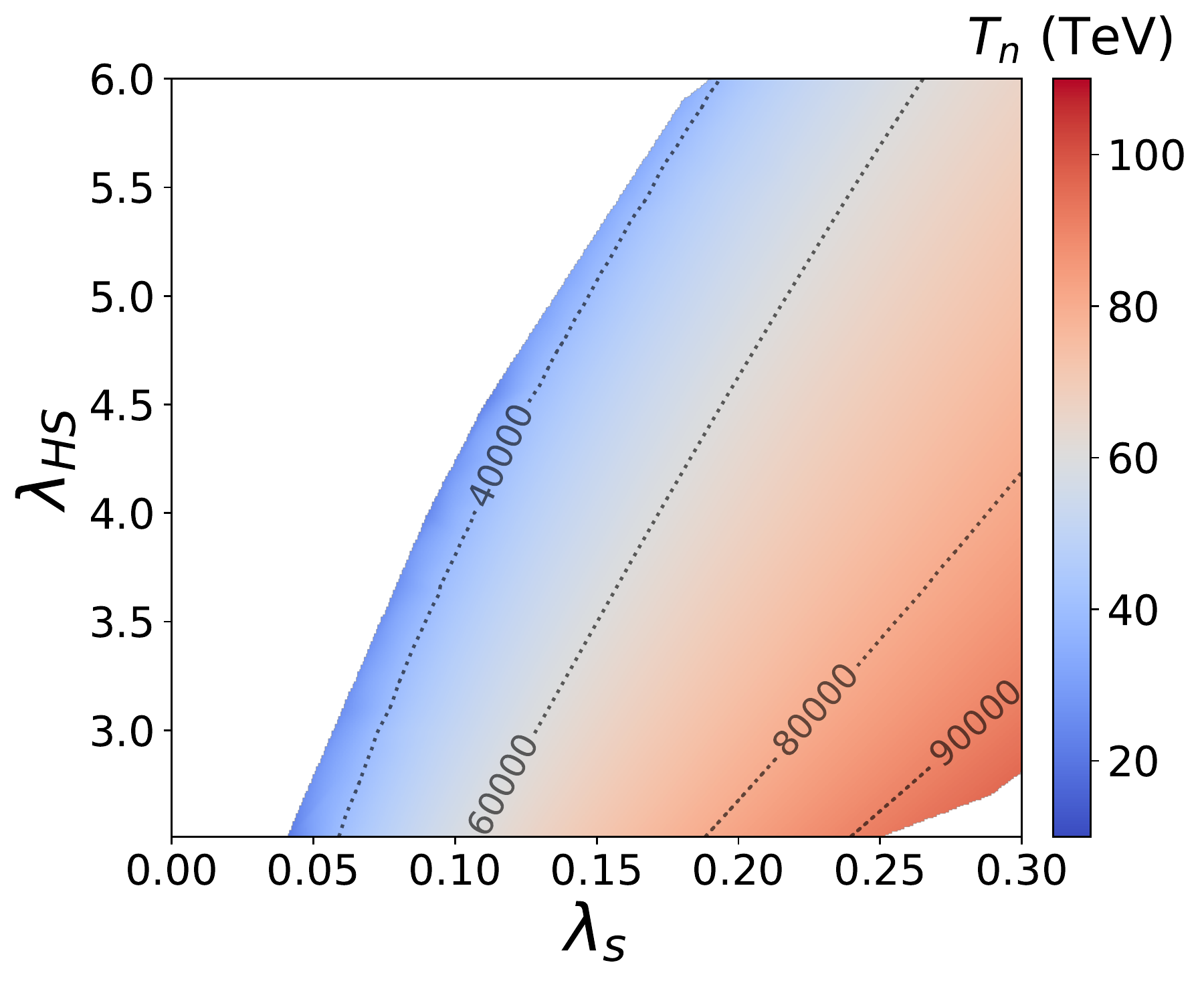}
\caption{\label{fig: param_scan2} PT parameters for model 2 with  flavon VEV, $v_s=10^5$ GeV, and $y=0.5$. $\alpha$ (left), $\beta/H_*$ (middle), $T_n$ (right). Left panel: variation of $\alpha$, with contours $\alpha=0.012,~0.02,~0.03,~0.04,~0.05$. Middle panel: $\beta/H_*$ is depicted with contours $\beta/H_*= 3\times10^2,~ 10^3,10^4,~10^5,~10^6$. Right panel: $T_n$ is depicted with contours $T_n=4,~6,~8,~9$ $\times 10^4$ GeV.}
\end{figure}

Scans showing the allowed region of FOPT for model 1 are given in figure \ref{fig: param_scan1}. We see the same qualitative trends as we move from left to right in the plots of $\alpha$, $\beta/H_*$ and $T_n$ as in case of the reference model. The biggest difference is seen in the left panel depicting $\alpha$, where the change near the left boundary is more dramatic compared to the reference model. The effect of the VLQ's of model 1 is therefore to reduce the region of SFOPT slightly. The color gradient in the plots of $\beta/H_*$ and $T_n$ is similarly, more dramatic as compared to that of the reference model. To note the differences between model 1 and model 2, we overlay contours for a few chosen values of $\alpha$, $\beta/H_*$ and $T_n$.

In figure \ref{fig: param_scan2}, we show the parameter scans for model 2, with the same choice of contours as model 1. In the left panel depicting $\alpha$, we see that the contrours for $\alpha=0.012,~0.02$ appear more spread out when compared to the respective contours for model 1, indicating a slower fall in the $\alpha$ values in model 2 as we increase $\lambda_S$. Close to the left boundary however, the contours of the two models are comparable. We find that the plots resemble those of the reference plots more closely as compared to model 1. This is because of the larger scale separation between $v_s$ and the VLQ mass scale $M$ for model 2, compared to model 1, due to which the VLQ contribution is more suppressed in model 2. As a result, the influence of VLQ's is less pronounced in model 2 as compared to model 1. In the next section, we describe the GW signature arising from a SFOPT.

\section{Gravitational wave background}
\label{sec:GW}
The GW spectrum is defined in terms of the GW energy density $\rho_{\rm{GW}}$, as \cite{Caprini:2015zlo},
\begin{equation}
\Omega_{\rm{GW}}(f)\equiv \frac{1}{\rho_c}\frac{d\rho_{\rm {GW}}}{d\ln f},
\end{equation}
where $f$ is the frequency and $\rho_c$ is the critical energy density of the universe, given by,
\begin{equation}
\rho_c = \frac{3H_0^2}{8\pi G}.
\end{equation}
\noindent Here, $H_0 = 100h~ {\rm{km/s/Mpc}}$, is the Hubble constant today, with the current value of $h=0.675\pm0.005$, and $G$ is the Newton's gravitational constant.

% If there is sufficient friction in the plasma, the bubble walls end up reaching a terminal velocity $v_w$. This is called the non-runaway scenario. On the on the other hand, if the friction provided by the plasma is not sufficient, the bubble walls may continue to accelerate forever in the plasma rest frame, which is referred to as the runaway case. 

Given a strong FOPT, bubbles of the stable phase larger than a critical size expand rapidly.  If there is sufficient friction in the plasma, the bubble walls may end up reaching a terminal velocity $v_w$. GW's are produced when the bubbles collide and coalesce with each other. The latent heat of the colliding bubbles gets distributed into GW's via three main processes: bubble wall collisions ($\Omega_{\rm{col}}$), sound waves produced in the thermal plasma ($\Omega_{\rm{sw}}$), and the resulting MHD turbulence ($\Omega_{\rm{turb}}$). The contributions can be added in the linear approximation:
\begin{equation}
h^2 \Omega_{\rm{GW}} \simeq h^2 \Omega_{\rm{col}}+h^2 \Omega_{\rm{sw}}+h^2 \Omega_{\rm{turb}}.
\end{equation}

In this paper, we have considered the spontaneous breaking of a global $U(1)_{\rm{FN}}$ symmetry, where no gauge bosons acquire a mass when they cross the bubble walls from the symmetric phase to the broken phase.  Also, there are no other light particles in the symmetric phase which become heavy in the broken phase. As a result, the friction exerted by the plasma on the bubble walls is expected to be negligible, which means that the walls can accelerate without any bounds. The runaway scenario \citep{Espinosa:2010hh,Bodeker:2017cim} is therefore relevant, and we take $v_w=1$. The contribution from bubble walls is expected to be the most significant while the effect of sound waves and MHD turbulence is subdominant. For the sake of completeness though, we list all three contributions to $h^2\Omega_{\rm{GW}}$ \cite{Caprini:2015zlo} in the runaway scenario , 
\begin{eqnarray}
h^2\Omega_{\rm{col}}(f) &=& 1.67\times 10^{-5} \left(\frac{H_*}{\beta}\right)^2 \left(\frac{\kappa_{\phi}\alpha}{1+\alpha}\right)^2\left(\frac{100}{g_*}\right)^{1/3}\left(\frac{0.11v_w^3}{0.42+v_w^2}\right)S_{\rm{col}}(f),\label{eq: collisions}\\
h^2\Omega_{\rm{sw}}(f) &=& 2.65\times 10^{-6} \left(\frac{H_*}{\beta}\right)^2 \left(\frac{\kappa_{v}\alpha}{1+\alpha}\right)^2\left(\frac{100}{g_*}\right)^{1/3}v_w ~S_{\rm{sw}}(f),\label{eq: soundwaves}\\
h^2\Omega_{\rm{turb}}(f) &= &3.35\times 10^{-4} \left(\frac{H_*}{\beta}\right)^2 \left(\frac{\kappa_{\rm{turb}}\alpha}{1+\alpha}\right)^{3/2}\left(\frac{100}{g_*}\right)^{1/3}v_w ~S_{\rm{turb}}(f). \label{eq: turbulence}
\end{eqnarray}

\noindent The spectral shape functions, $S_{\rm{col}}$, $S_{\rm{sw}}$ and $S_{\rm{turb}}$ determine the power law behaviour of each contribution at low and high frequencies. These are, 
\begin{eqnarray}
S_{\rm{col}}(f) &=& \left(\frac{f}{f_{\rm{col}}}\right)^{2.8}\frac{3.8}{1+2.8(f/f_{\rm{col}})^{3.8}},\\
S_{\rm{sw}}(f) &=& \left(\frac{f}{f_{\rm{sw}}}\right)^3 \left(\frac{7}{4+3(f/f_{\rm{sw}})^2}\right)^{7/2},\\
S_{\rm{turb}}(f) &=& \left(\frac{f}{f_{\rm{turb}}}\right)^3\frac{1}{[1+(f/f_{\rm{turb}})]^{11/3}(1+8\pi f/h_*)}.
\end{eqnarray}

Here, $h_*$ is the Hubble rate at $T=T_*$, 
\begin{equation}
h_* = 1.65\times 10^{-7}~{\rm{Hz}}\left(\frac{T_*}{100~{\rm{GeV}}}\right)\left(\frac{g_*}{100}\right)^{1/6}.
\end{equation} 
Taking into account the expansion of the universe from $T=T_*$ to present day, the red-shifted peak frequencies are,
\begin{eqnarray}
f_{\rm{col}} &=& 1.65\times 10^{-5}~{\rm{Hz}}
\left(\frac{\beta}{H_*}\right)\left(\frac{T_*}{100~{\rm{GeV}}}\right)\left(\frac{g_*}{100}\right)^{1/6}\left(\frac{0.62}{1.8-0.1v_w+v_w^2}\right),\label{eq: fcol}\\
f_{\rm{sw}} &=& 1.9\times 10^{-5}~{\rm{Hz}}~\frac{1}{v_w}\left(\frac{\beta}{H_*}\right)\left(\frac{T_*}{100~{\rm{GeV}}}\right)\left(\frac{g_*}{100}\right)^{1/6},\label{eq: fsw}\\
f_{\rm{turb}} &=& 2.7\times 10^{-5}~{\rm{Hz}}~\frac{1}{v_w}\left(\frac{\beta}{H_*}\right)\left(\frac{T_*}{100~{\rm{GeV}}}\right)\left(\frac{g_*}{100}\right)^{1/6}.\label{eq: fturb}
\end{eqnarray}

We also need the efficiency factors: $\kappa_{\phi}$, $\kappa_{\rm{sw}}$, $\kappa_{\rm{turb}}$ representing the fraction of the latent heat which is shared by the three different processes. In the runaway scenario, these are,
\begin{equation}
\kappa_{\phi} = \frac{\alpha-\alpha_{\infty}}{\alpha}, ~~\kappa_v = \frac{\alpha_{\infty}}{\alpha}\kappa_{\infty}, ~~\kappa_{\rm{turb}} = \epsilon_{\rm{MHD}}\kappa_v.
\end{equation}
Here, $\epsilon_{\rm{MHD}}$ is the turbulent fraction of bulk motion in plasma, which is at most $5\%$-$10\%$ \cite{Hindmarsh:2015qta}. Also, $\kappa_{\infty}$ is given by,
\begin{eqnarray}
\kappa_{\infty} &\equiv& \frac{\alpha_{\infty}}{0.73+0.083\sqrt{\alpha_{\infty}}+\alpha_{\infty}}, \\
\alpha_{\infty} &\simeq& \frac{5}{4\pi^2}\frac{\sum_i c_i\Delta m_i^2}{g_*T_*^2},
\end{eqnarray}
where the sum over $i$ runs over all particles that are light in the symmetric phase and acquire mass in the broken phase, with $\Delta m_i^2$ as the mass squared difference in the two phases. Also,
$c_i = n_i (n_i/2)$ for bosons (fermions) with $n_i$ the number of degrees of freedom of the particle. Since in our case there are no such light particles which become heavy as they cross the bubble wall, we expect $\alpha_{\infty}$ to be negligible. As a result $\alpha>>\alpha_{\infty}$, and only the contribution from bubble collisions is significant,
\begin{equation}
h^2\Omega_{\rm{GW}} \simeq h^2\Omega_{\rm{col}}.
\end{equation}

\subsection{Detection prospects}
To get a detectable GW signal resulting from FOPT, we need large $\alpha$, and small $\beta/H_*$, as is evident from the dependence of expressions \eqref{eq: collisions}, \eqref{eq: soundwaves} and \eqref{eq: turbulence} on these quantities. In the parameter scans considered in section \ref{sec:FOPT}, we saw that the strongest FOPT occurs near the left boundary as shown in figure \ref{fig: noVLQ}, figure \ref{fig: param_scan1}, and figure \ref{fig: param_scan2}.

\begin{figure}[tbp]
\centering % \begin{center}/\end{center} takes some additional vertical space
\includegraphics[width=.49\textwidth]{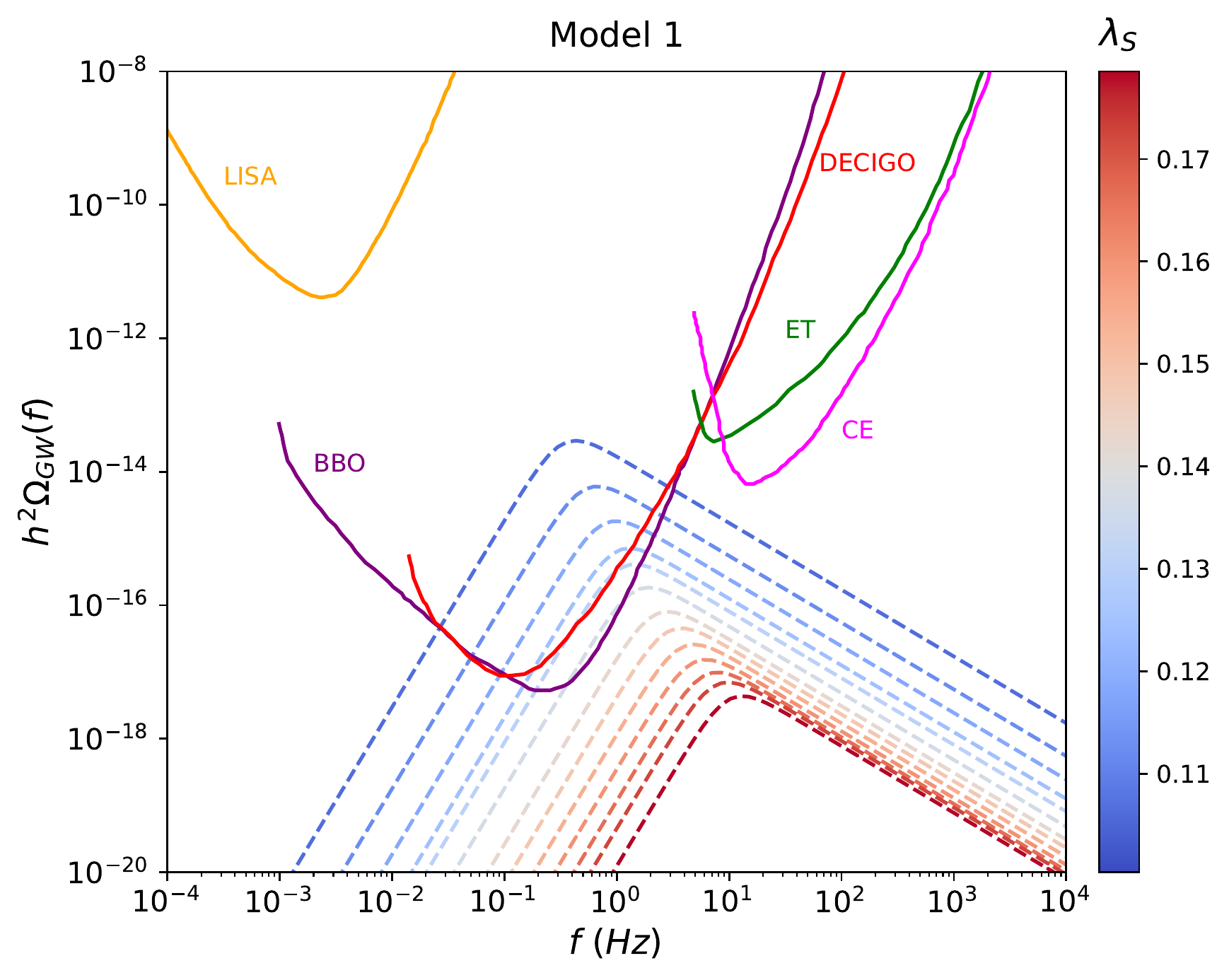}
\hfill
\includegraphics[width=.49\textwidth]{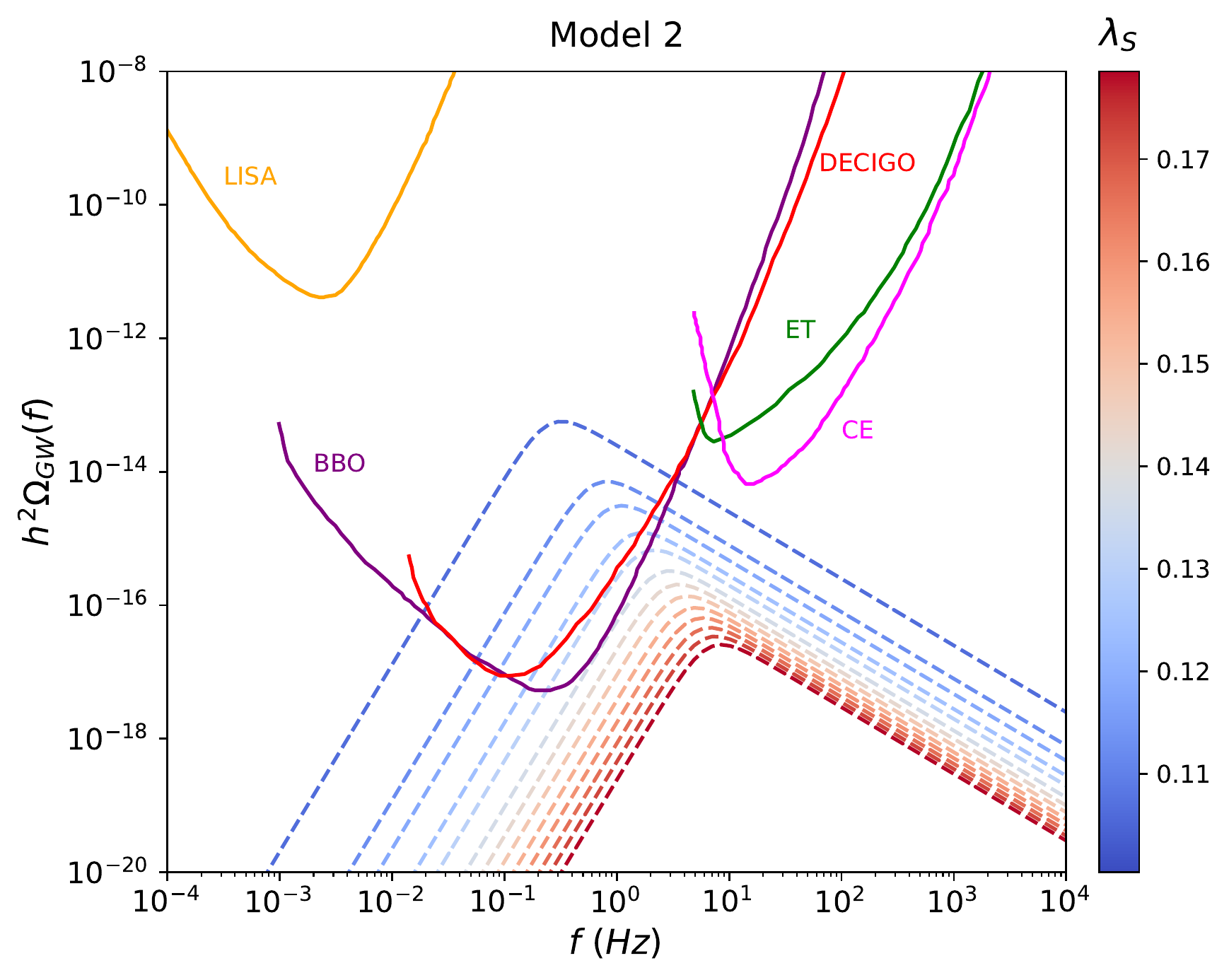}
\caption{\label{fig: gw} GW spectra for model 1 and model 2 for $\lambda_s\in[0.1,0.18]$. Here $\lambda_{HS} = 4.253$, $v_s=10^5$ GeV, and $y=0.5$.}
\end{figure}

\begin{figure}[tbp]
\centering % \begin{center}/\end{center} takes some additional vertical space
\includegraphics[width=.47\textwidth]{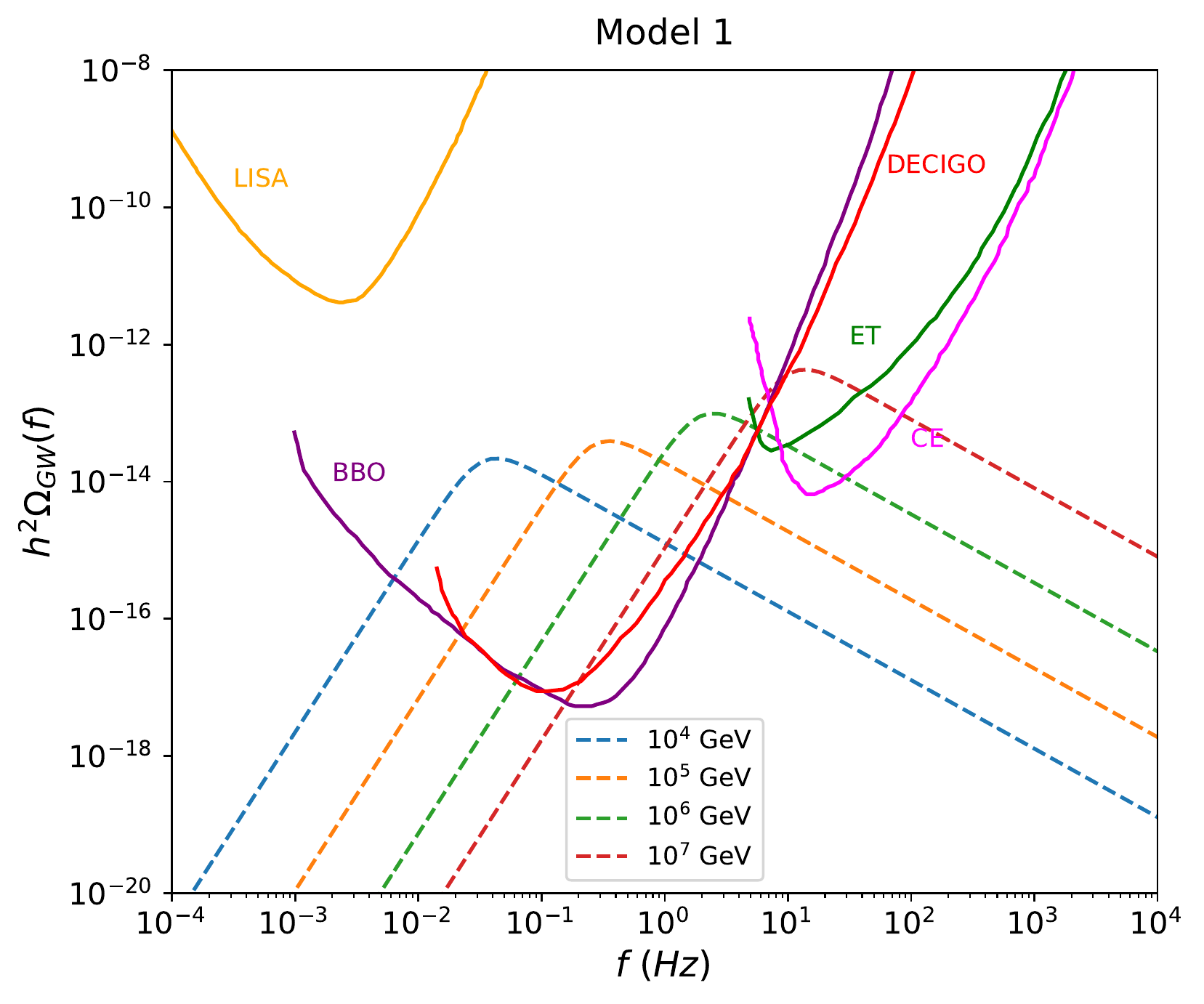}
\hfill
\includegraphics[width=.47\textwidth]{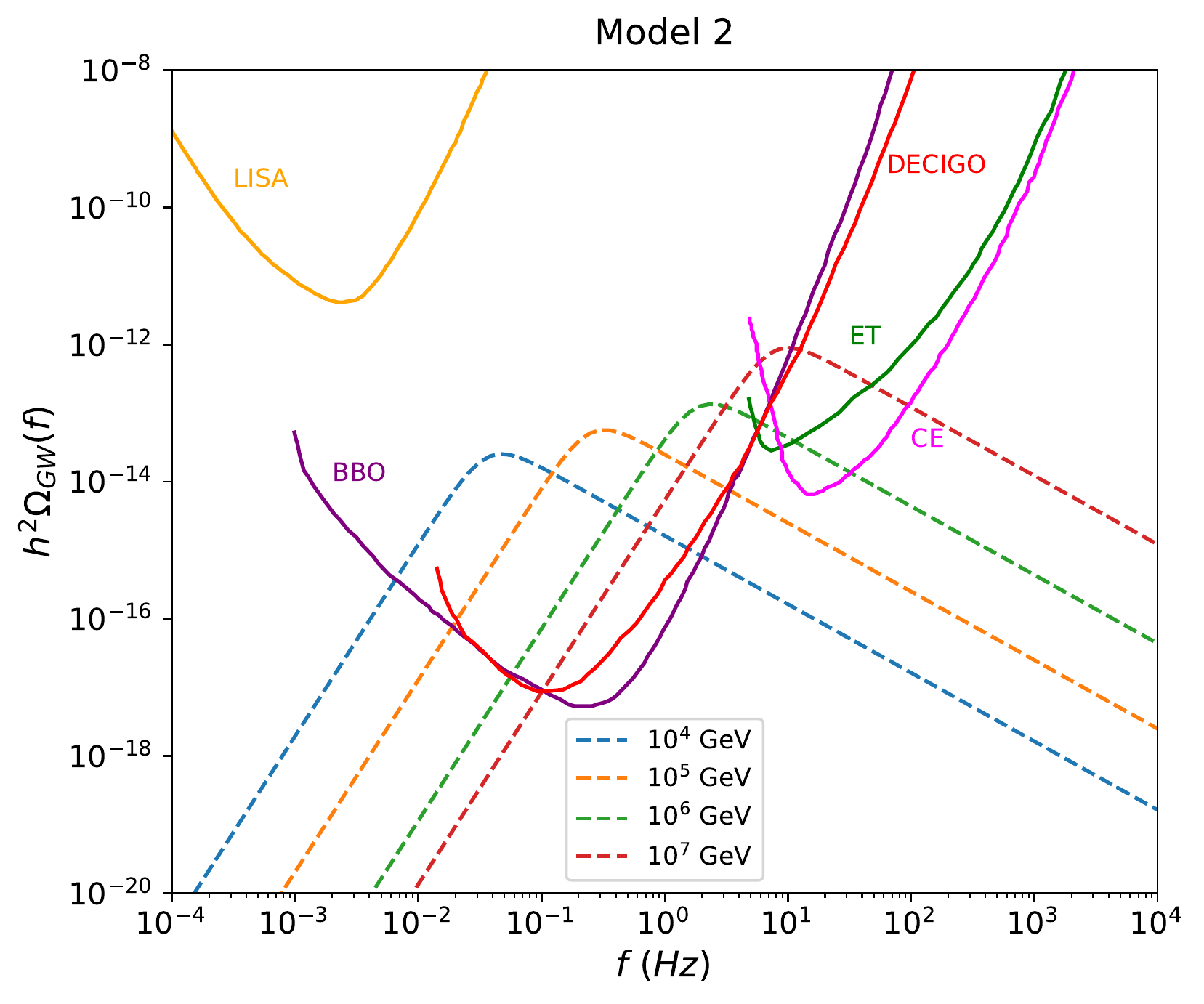}
\caption{\label{fig: gw_vs} GW spectra for model 1 and model 2 for $v_s = 10^{4,5,6,7}~{\rm{GeV}}$. Here $\lambda_s =0.105$, $\lambda_{HS} = 4.253$, and $y=0.5$. The peak frequency is seen to shift from left to right as $v_s$ increases.}
\end{figure} 

An example of GW spectra highlighting the dependence on model parameters, is shown in figure \ref{fig: gw}. For model 1 and model 2, the dashed curves show the GW spectrum, when $\lambda_{HS} = 4.253$, and $\lambda_S$ is varied from $\lambda_S=0.1$ to $\lambda_S$ =0.18. The other parameters are kept fixed at $v_s=10^5$ GeV, and $y=0.5$. Solid curves are the power law integrated sensitivity curves (PLISC) \cite{Thrane:2013oya} for the GW observatories LISA \cite{LISA:2017pwj}, BBO \cite{Corbin_2006}, DECIGO \cite{Musha:2017usi}, ET \cite{Punturo:2010zz} and CE \cite{Abbott_2017}. The strength of the GW spectrum decreases as $\lambda_S$ increases. This corresponds to moving horizontally from left to right, starting close to the boundary  in the $\lambda_S$-$\lambda_{HS}$ plane in figures \ref{fig: param_scan1} and \ref{fig: param_scan2}. As the GW strength decreases, it eventually goes below the noise floor of DECIGO and BBO, becoming undetectable.

In figure \ref{fig: gw_vs}, we show the effect of varying the scale $v_s$  on the GW spectrum for a benchmark value of $(\lambda_S,\lambda_{HS})$. Taking $v_s = 10^4,~10^5, ~10^6,~10^7~{\rm{GeV}}$, we observe that the peak frequency shifts from left to right as $v_s$ increases. This can be explained from \eqref{eq: fcol},\eqref{eq: fsw} and \eqref{eq: fturb}, where the peak frequency is proportional to $T_*\sim v_s$. The height of the peak amplitude is seen to increase with $v_s$ due to the temperature dependence of the RHS of \eqref{eq: nuclCriterion}, which leads to a larger amount of supercooling at higher $T$. Clearly, different detectors are suitable for different scales $v_s$. For $v_s =10^{4,5}$ GeV, the chosen benchmark point gives a strong GW signal at DECIGO and BBO for both model 1 and model 2. While for $v_s=10^6$ GeV, the signal would also be detectable at CE, in addition to DECIGO and BBO. For $v_s=10^7$ GeV, the signal can be detected at BBO, DECIGO, CE and ET. For the entire viable region of FOPT, the GW signal is not observable at LISA, since the peak amplitude is almost always found to be smaller than $10^{-12}$.

\begin{figure}[tbp]
\centering % \begin{center}/\end{center} takes some additional vertical space
\includegraphics[width=.47\textwidth]{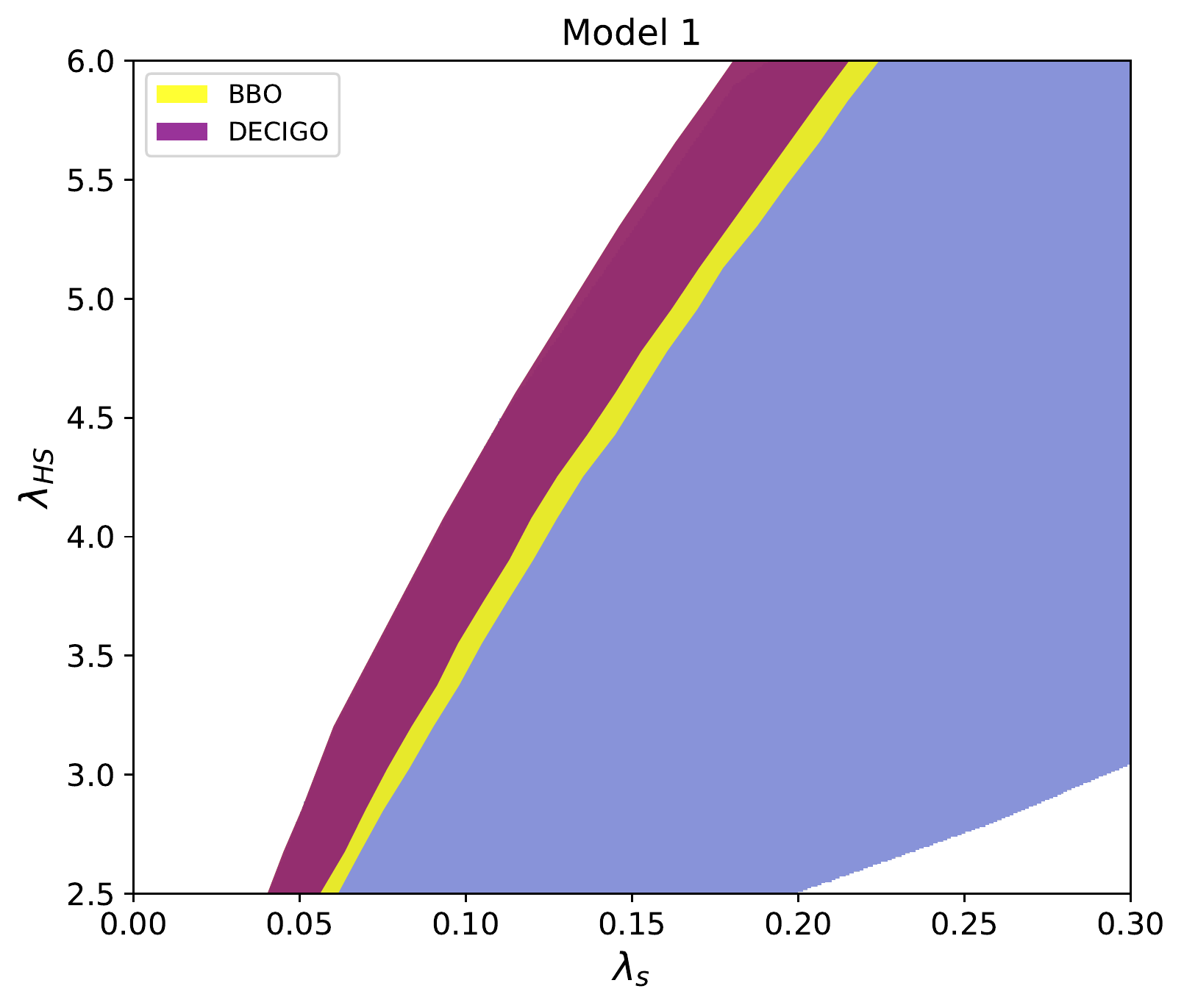}
\hfill
\includegraphics[width=.47\textwidth]{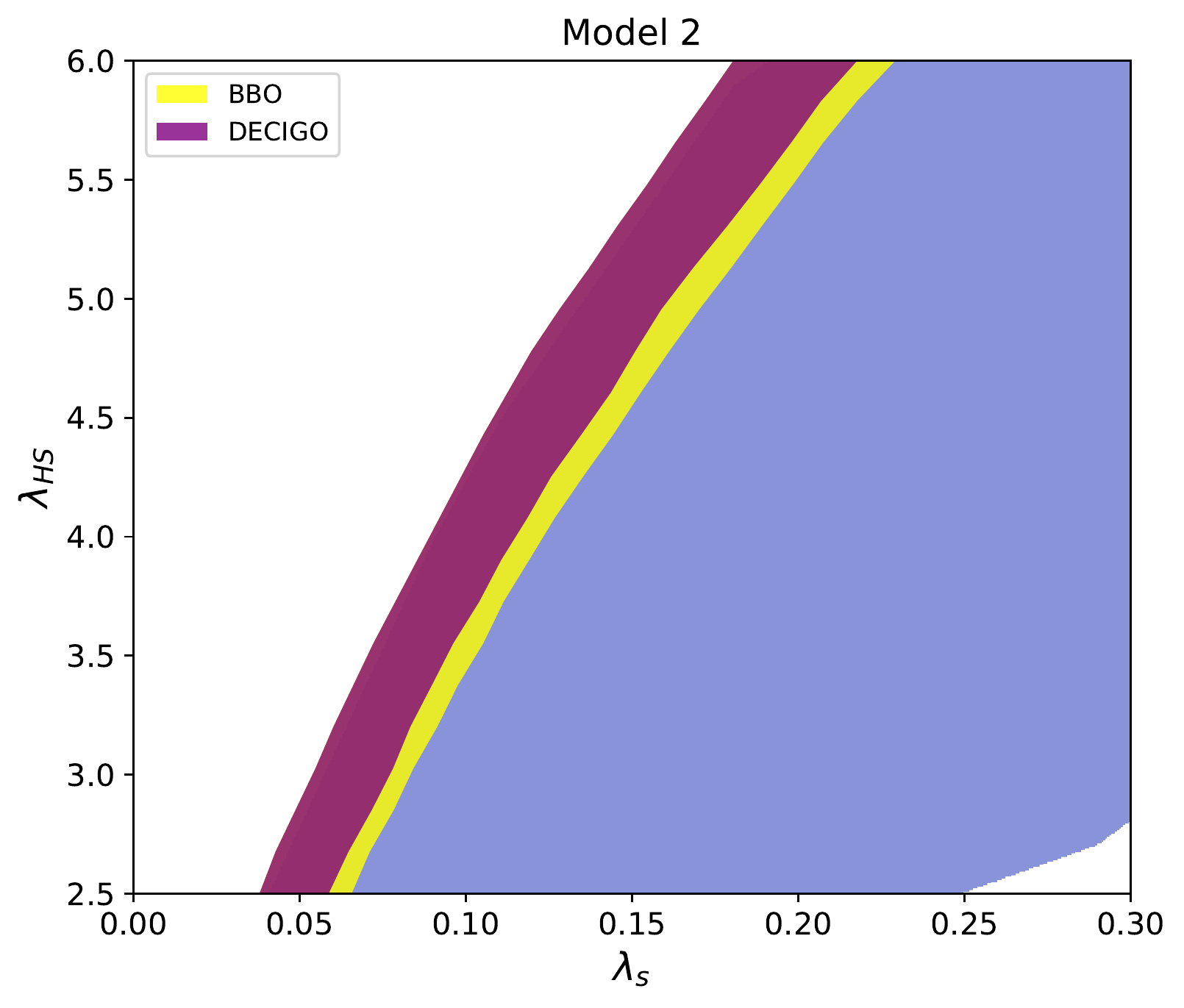}
\caption{\label{fig: gw_detection} Regions of GW detection for model 1 and model 2, with $v_s=10^5$ GeV, and $y=0.5$. The region coloured in purple overlaps the yellow region, since BBO is more sensitive compared to DECIGO.} 
\end{figure}

\begin{figure}[tbp]
\centering % \begin{center}/\end{center} takes some additional vertical space
\includegraphics[width=.47\textwidth]{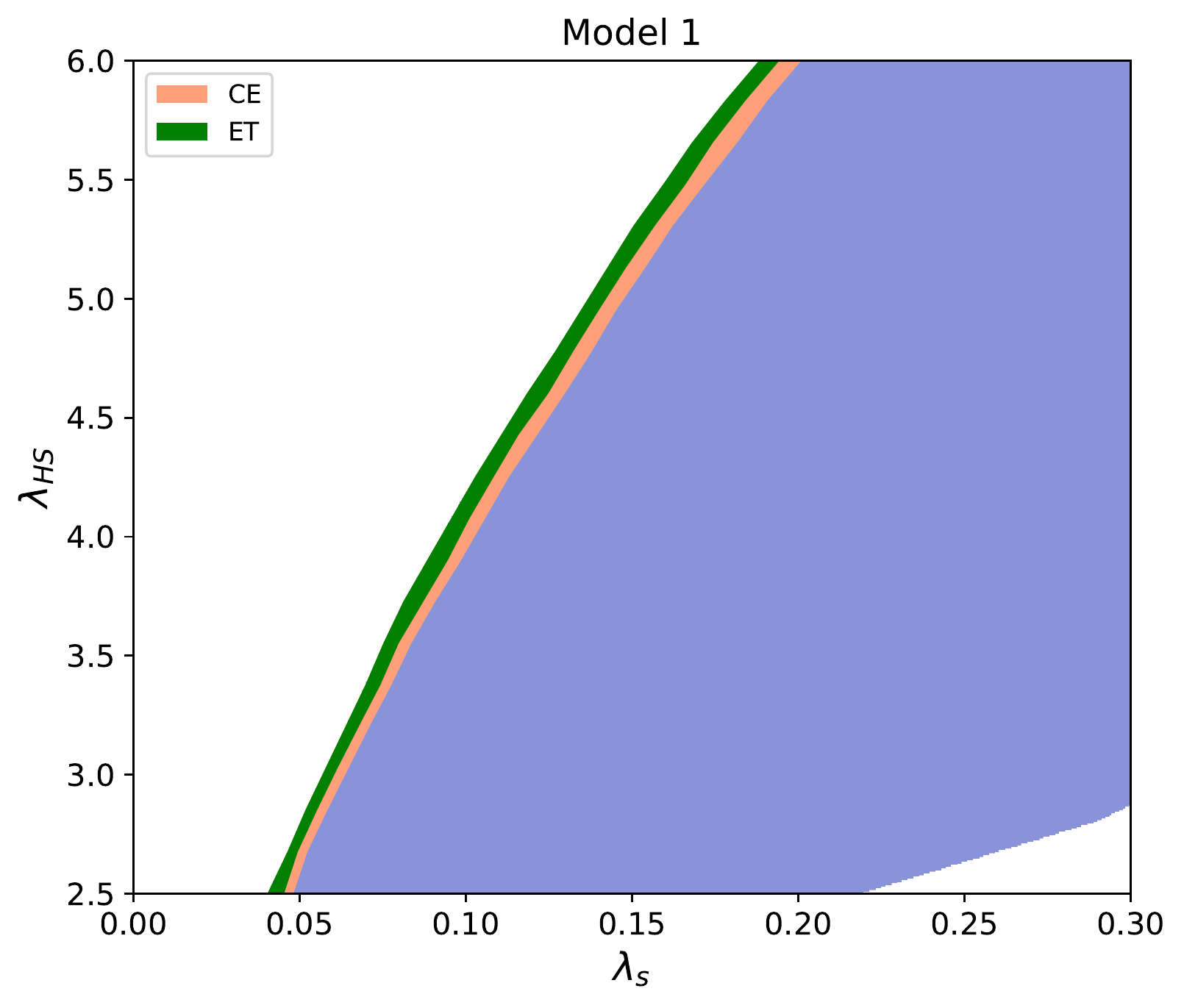}
\hfill
\includegraphics[width=.47\textwidth]{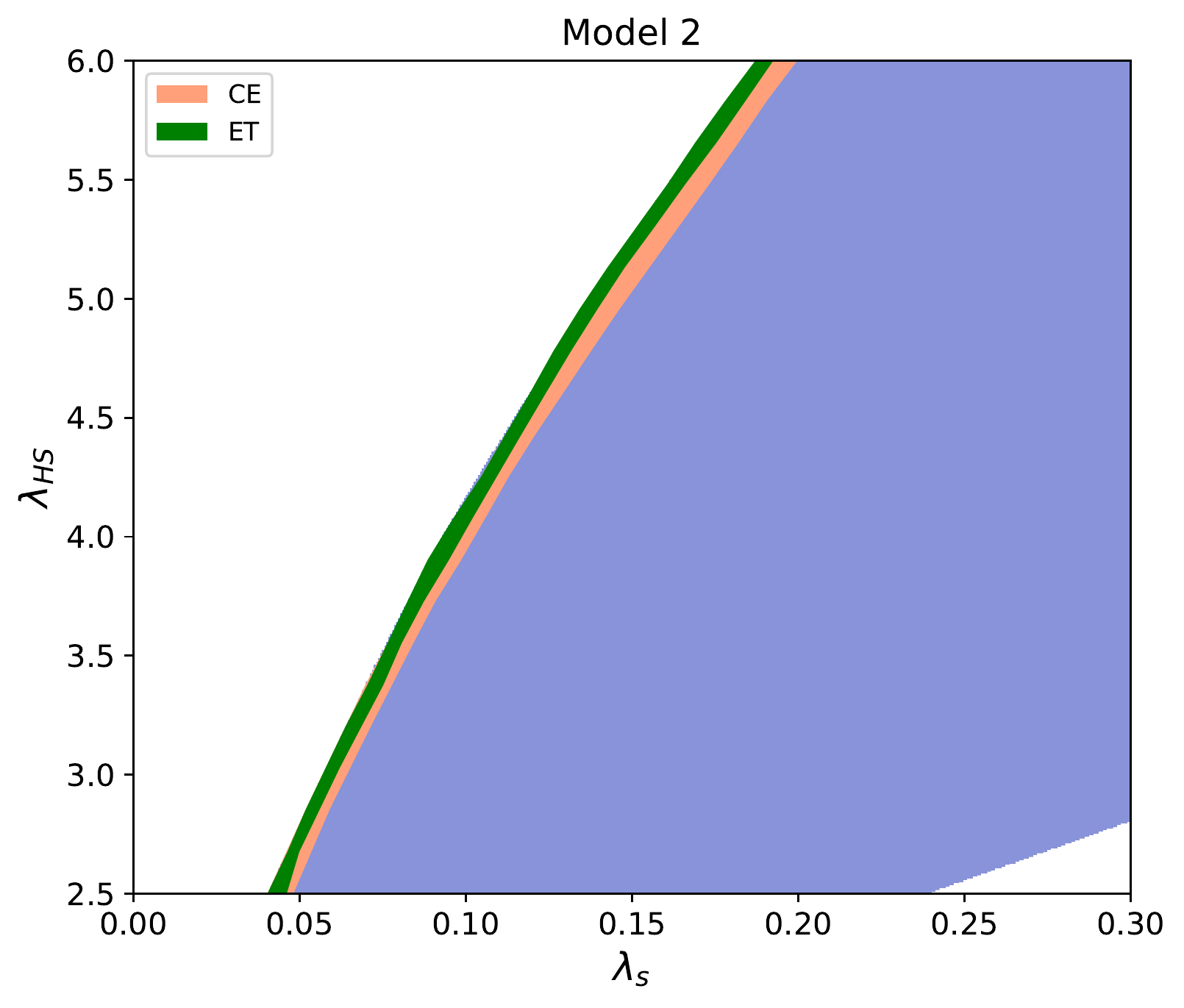}
\caption{\label{fig: gw_detection_vs1.e7} Regions of GW detection for model 1 and model 2, with $v_s=10^7$ GeV, and $y=0.5$. The region coloured in green overlaps the peach region, since CE is more sensitive compared to ET.} 
\end{figure}

Next, in figure \ref{fig: gw_detection}, we showcase the detectable regions of the parameter space for the model 1 (left panel) and model 2 (right panel), when $v_s=10^5$ GeV, and $y=0.5$. The  coloured region corresponds to the region sensitive to DECIGO (purple), and BBO (yellow). The two regions overlap each other; the region corresponding to BBO is broader than that of DECIGO, since the former is more sensitive. In the blue region, the GW signal is too weak to be detected. In figure \ref{fig: gw_detection_vs1.e7}, we show the detectable parameter space for both models with $v_s=10^7$ GeV, and $y=0.5$. Now the parameter space is accessible to ET and CE in addition to DECIGO and BBO. For convenience we show only regions for ET (green) and CE (peach), as the regions for BBO and DECIGO are much smaller. The detectable region is narrower compared to the region at $10^5$ GeV.  The best detection prospects for $v_s=10^7$ GeV are at CE.

Previously in figure \ref{fig: param_scan1} and figure \ref{fig: param_scan2}, it was found that the constant $\alpha$ contours for chosen values lie closer to the boundary for model 1 as compared to model 2. However it is interesting to note that apart from minor qualitative differences, the GW signal strength is comparable for the two models in the detectable region, both at $v_s=10^5$ GeV, and $v_s=10^7$ GeV. This is because the regions of detectability require $\alpha\gtrsim 0.03$, where the contours in the two models are similar. This implies that a detectable GW background cannot discrimitate between model 1 and model 2. The GW signal is therefore not sensitive to the specific $U(1)_{\fn}$ charge assignment, as long as there is only one flavon which breaks the symmetry. 

\section{Conclusions}\label{sec: conclusion}
In this paper, we have explored the possibility of probing FN models with GW detectors, if the flavon undergoes a SFOPT. To do so, we have constructed two minimal, non-supersymmetric UV completions of the FN mechanism involving a single flavon.  To avoid any possible complications arising from gauging it, we have chosen the family symmetry group as global $U(1)$. However, the two models considered here are UV complete only in the sense that they generate the effective FN operators which give rise to the hierarchy of quark yukawas at low scale. We find that it is essential to add heavy bosons couple to the flavon, to counter the destabilizing effect of adding several fermions to the theory. We have chosen a region of parameter space where the effect of the heavy bosons on the nature and strength of FOPT is insignificant, to make  predictions independent of any specific realization of heavy bosons. 

To avoid stringent constraints from collider experiments, we have focussed on the intermediate energy scale, $10^4$-$10^7$ GeV for the scale of the FN mechanism. We find that in the $\lambda_{S}$-$\lambda_{HS}$ plane, a strong FOPT is observed near the boundary separating the FOPT region from the region where the PT never completes. The region for observing a GW signal therfore lies within a thin strip close to this boundary. Detection prospects are the best when $v_s \sim 10^4$-$10^5$ GeV, in which case the peak of the GW signal lies in the sensitivity range of BBO and DECIGO. Higher scales of $v_s \sim 10^6$-$10^7$ GeV can be probed by BBO, DECIGO, as well as CE and ET.

Detecting a stochastic GW background can become a powerful tool to probe the flavor symmetry-breaking scale in the coming years, along with complementary collider signatures. For the two models considered, the GW signatures are of comparable strength when the same set of parameters is used. Therefore we conclude that it is not possible to discriminate between them, based solely on the GW signature. We take the liberty to extrapolate this finding to all models which involve a single flavon, when the family symmetry group is $U(1)_{\fn}$. As such, models involving more than one flavon may be discriminated by GW's since they may feature multiple peaks, owing to different scales of the VEV's, but this possibility remains to be seen. It would  be interesting to see if the GW signature can throw some light on the FN symmetry group. 

\acknowledgments
D.R. would like to thank Subhendu Rakshit, Siddhartha Karmakar, Dipankar Das, and Najimuddin Khan for useful discussions. This work is supported by the Department of Science and Technology, India via SERB grant EMR/2014/001177 and DST-DAAD grant INT/FRG/DAAD/P-22/2018. D.R. is supported by the Govt. of India UGC-SRF fellowship.

\appendix

\section{A two-flavon model}\label{app: model3}
Consider the horizontal group: $G_{\fn} = U(1)_{\rm{H1}}\times U(1)_{\rm{H2}}$. The flavor symmery is then broken by two flavons $S_1,~S_2$, one for each $U(1)$ group, which get a VEV. A possible two-flavon model is given below \cite{Nir-Seiberg1}:

\begin{itemize}
\item \textbf{Model 3}: $\epsilon_1 \sim \lambda^2 = 0.04,~\epsilon_2\sim \lambda^3 = 0.008$,
$$Q_{\fn}(\overline{Q}_L)=((0,1),(1,0),(0,0)),$$
$$Q_{\fn}(u_R)=((0,1),(-1,1),(0,0)),$$
$$Q_{\fn}(d_R)=((0,1),(1,0),(1,0))$$
The yukawa matrices are:
\begin{gather}
{\bf{Y^u}} \sim \begin{pmatrix}
\epsilon_2^2 & \epsilon_1^{-1}\epsilon_2^2 & \epsilon_2\\
\epsilon_1\epsilon_2 & \epsilon_2 & \epsilon_1\\
\epsilon_2 & \epsilon_1^{-1}\epsilon_2 & 1
\end{pmatrix},~~~
{\bf{Y^d}} \sim \begin{pmatrix}
\epsilon_2^2 & \epsilon_1\epsilon_2 & \epsilon_1\epsilon_2\\
\epsilon_1\epsilon_2 & \epsilon_1^2 & \epsilon_1^2\\
\epsilon_2 & \epsilon_1 & \epsilon_1
\end{pmatrix} .
\end{gather}

The determinants are:
\begin{equation}\label{det3}
\det {\bf{Y^u}} \sim \epsilon_2^3,~~~\det {\bf{Y^d}} \sim \epsilon_2^2\epsilon_1^3
\end{equation}
\end{itemize}
In supersymmetric models, negative powers of $\epsilon$ are prohibited due to holomorphy; this gives rise to texture zeros (see \cite{Nir-Seiberg1}). In this work, we have only considered non-supersymmetric realizations of FN. 

\section{Stabilizing the potential with heavy bosons}\label{app: bosons}
Here we illustrate a scenario under which the effective potential can be stabilized by heavy bosons. Consider $N$ bosons at mass scale $M_b$, with an unbroken $O(N)$ symmetry:
\begin{eqnarray}
\mathcal{L} &\supset & \frac{1}{2}\big(\partial^{\mu}\Phi\big)^{\dagger}\big(\partial_{\mu}\Phi\big) - V(\Phi,S)\nonumber\\
V(\Phi,S) &=& \frac{1}{2}M_b^2\big(\Phi^{\dagger}\Phi\big) + \frac{1}{4}\lambda_b\big(\Phi^{\dagger}\Phi\big)^2 + \frac{1}{2}\lambda_{bS} \big(\Phi^{\dagger}\Phi\big)|S|^2,
\end{eqnarray} 
where $\Phi$ is a column vector of $N$ real scalar fields, and $\lambda_b$ and $\lambda_{bS}$ are the new couplings. Including the Coleman-Weinberg correction from these bosons modifies $V_{\rm{CW}}$ as,

\begin{equation}\label{eq: CWboson}
V_{\rm{CW}} = \frac{1}{64\pi^2}\sum_{i} (-1)^{f_i}n_i m^4_i\bigg[\log\bigg(\frac{m^2_i}{\mu^2}\bigg)- c_i\bigg] + \frac{1}{64\pi^2}\sum_{b} m^4_i\bigg[\log\bigg(\frac{m^2_b}{\mu^2}\bigg)- c_b\bigg],
\end{equation}
where the sum over $i$ includes SM fields, VLQ's and the flavon, while sum over $b$ includes the new heavy bosons whose field dependent mass $m^2_b$ is given by,
\begin{equation}
m^2_b(s) = M_b^2+\frac{1}{2}\lambda_{bS} s^2.
\end{equation}

We take $\mu^2=v_s^2$, implying $m_b^2>>\mu^2$, and hence the heavy bosons can be integrated out, with their effect captured by higher dimensional operators (see for example \cite{Blum:2015rpa}). Keeping the leading order EFT operator, \eqref{eq: CWboson} becomes, 

\begin{equation}\label{eq: CWboson}
V_{\rm{CW}} = \frac{1}{64\pi^2}\sum_{i} (-1)^{f_i}n_i m^4_i\bigg[\log\bigg(\frac{m^2_i}{\mu^2}\bigg)- c_i\bigg] + \frac{c_6}{\Lambda_b^2}s^6,
\end{equation}
with $c_6$ as the Wilson coefficient.

\bibliographystyle{unsrt}
\bibliography{citation.bib}

\end{document}